# Capability-Priced Micro-Markets: A Micro-Economic Framework for the Agentic Web over HTTP 402

Ken Huang, CSA and DistributedApps.ai
Jerry Huang, Kleiner Perkins
Mahesh Lambe, Unify Dynamics
Hammad Atta  Qorvex Consulting /Roshan Consulting
Dr.Yasir Mehmood Qorvex Consulting
Dr. Muhammad Zeeshan Baig Qorvex Consulting
Dr.MUHAMMAD AZIZ UL HAQ Qorvex Consulting
Dr.Muhammad Aatif Qorvex Consulting
Nadeem Shahzad , Roshan Consulting
Shailja Gupta, Carnegie Mellon University
Rajesh Ranjan, Carnegie Mellon University
Rekha Singhal, Tata Consultancy Services Research

**Abstract**

This paper introduces Capability-Priced Micro-Markets (CPMM), a micro-economic framework designed to enhance secure, scalable commerce among autonomous AI agents within the agentic web. Addressing the challenge of economic coordination in decentralized ecosystems, CPMM enables transactions with minimal human oversight by integrating three foundational technologies: the Project NANDA infrastructure for cryptographically verifiable, capability-based security and discovery; the HTTP 402 "Payment Required" status code, enriched with contemporary X402/H402 extensions to facilitate efficient micropayments; and the Agent Capability Negotiation and Binding Protocol (ACNBP) for secure, multi-step negotiation and commitment. We model agent interactions as a repeated bilateral game with incomplete information, proving that the CPMM mechanism converges to a constrained Radner equilibrium, which guarantees efficient outcomes in the face of information asymmetry. A novel theoretical contribution is the concept of "privacy elasticity of demand," which quantifies the trade-off between an agent's information disclosure and the market price of its services. By merging secure capabilities, micropayment protocols, and formal negotiation mechanisms, CPMM provides a comprehensive, theoretically robust solution for the development of functional micro-markets in the evolving landscape of the agentic web. This framework paves the way for more effective economic interactions among autonomous agents, fostering innovation and efficiency in digital commerce.

# 1. Introduction

Agentic Web represents a paradigm shift in distributed computing, where specialized AI entities must coordinate, negotiate, and transact with minimal human intervention. As these agent ecosystems proliferate across domains ranging from financial services to scientific research, a fundamental bottleneck has emerged: the lack of robust economic coordination mechanisms that can operate at the scale, speed, and granularity required by autonomous systems.

Traditional approaches to agent coordination have relied on either centralized marketplaces with human oversight or simplified protocols that assume static pricing and homogeneous agent populations. However, the reality of modern agent ecosystems is far more complex. Agents possess heterogeneous capabilities that may be context-dependent, dynamically priced, and subject to real-time availability constraints. Furthermore, the transient nature of many agent interactions—often lasting mere seconds or minutes—demands economic mechanisms that can operate with minimal latency while maintaining security and verifiability.

The challenge is not merely computational but fundamentally economic: how does an autonomous agent determine what to charge for its services? How does a buyer agent discover and evaluate potential service providers in a decentralized network? How can payments be settled when transaction values may be measured in fractions of cents, and how can trust be established between pseudonymous entities that may never interact again?

Existing protocols in the agent communication landscape, including the Model Context Protocol (MCP) [1], Agent-to-Agent (A2A) frameworks [2], and Cisco's Agent Gateway Protocol (AGP) [3], have made significant strides in enabling agent interoperability. However, these protocols either assume static pricing models or delegate economic coordination to external systems, leaving a critical gap in the infrastructure needed for truly autonomous economic agents.

This paper introduces Capability-Priced Micro-Markets (CPMM), a unified economic framework that addresses these challenges by integrating three complementary technological advances into a coherent system for autonomous agent commerce. The framework builds upon MIT's NANDA(Networked Agents and Decentralized AI) infrastructure [4], which provides capability-based security and federated agent discovery; HTTP 402 and its modern successors X402/H402 [5], which enable sub-cent micropayments with minimal transaction overhead; and the Agent Capability Negotiation and Binding Protocol (ACNBP) [6], which provides secure, verifiable negotiation and commitment mechanisms.

## 1.1 The Economic Coordination Problem

The fundamental challenge in autonomous agent economics lies in the intersection of several complex problems. First, the **price discovery problem**: in a decentralized network where agents have private information about their costs and capabilities, how can market-clearing prices emerge without centralized coordination? Traditional auction mechanisms assume the presence of auctioneers and well-defined market sessions, assumptions that break down in the fluid, always-on world of agent interactions.

Second, the **capability verification problem**: unlike traditional e-commerce where products have standardized descriptions, agent capabilities are often complex, context-dependent, and difficult to verify ex-ante. An agent claiming to provide "natural language translation" might excel at technical documents but fail on colloquial text, or perform well in low-latency scenarios but degrade under load. The economic mechanism must account for this uncertainty while providing incentives for honest capability advertisement.

Third, the **micropayment feasibility problem**: many agent interactions involve small-value transactions that would be economically infeasible under traditional payment systems due to fixed transaction costs. A language model inference might be worth $0.001, but processing this payment through conventional channels could cost $0.30 in fees. This creates a fundamental mismatch between the granularity of value creation and the economics of value transfer.

Fourth, the **trust and reputation problem**: in a pseudonymous network where agents may be ephemeral and interactions are brief, traditional reputation systems based on long-term relationships become inadequate. The economic mechanism must enable trust formation and maintenance in an environment where identity is fluid and history may be limited.

## 1.2 The CPMM Solution Architecture

CPMM addresses these challenges through a layered architecture that combines cryptographic capabilities, economic incentives, and protocol design. At its foundation, the system leverages MIT NANDA's capability-based security model, where agents possess cryptographically verifiable capabilities that serve both as authorization tokens and as the basis for economic valuation.

The NANDAinfrastructure provides several critical components for CPMM. Its federated Agent Name Service (ANS) enables efficient discovery of potential trading partners while preserving privacy through selective capability disclosure. The capability-based security model ensures that agents can only advertise services they can actually provide, reducing the scope for fraudulent capability claims. Most importantly, NANDA's attestation mechanisms provide cryptographic proof of capability provenance, enabling other agents to verify not just what capabilities an agent claims to possess, but the source and validity of those claims.

Building on this foundation, CPMM integrates HTTP 402 and its modern successors X402/H402 to enable efficient micropayments. The original HTTP 402 "Payment Required" status code was designed in the early days of the web to support digital cash systems that never materialized. However, recent developments in cryptographic payment systems have made the vision of frictionless micropayments technically feasible. The X402/H402 protocols extend HTTP 402 with modern cryptographic primitives, including ephemeral Ed25519 keys for transaction signing, Schnorr signatures for compact verification, and header-based payment metadata that enables rich payment semantics without requiring protocol modifications.

The third layer of CPMM is provided by ACNBP, which offers a structured 10-step protocol for capability negotiation and binding. ACNBP's innovation lies in its formal approach to the negotiation process, providing cryptographic guarantees about the integrity and non-repudiation of negotiation outcomes. The protocol's extension mechanism enables backward compatibility while allowing for protocol evolution, ensuring that CPMM can adapt to changing requirements without fragmenting the agent ecosystem. Figure 1 High Level CPMM stack overview shows how NANDA, ACNBP, and HTTP 402/X402/H402 compose into CPMM's layered system.

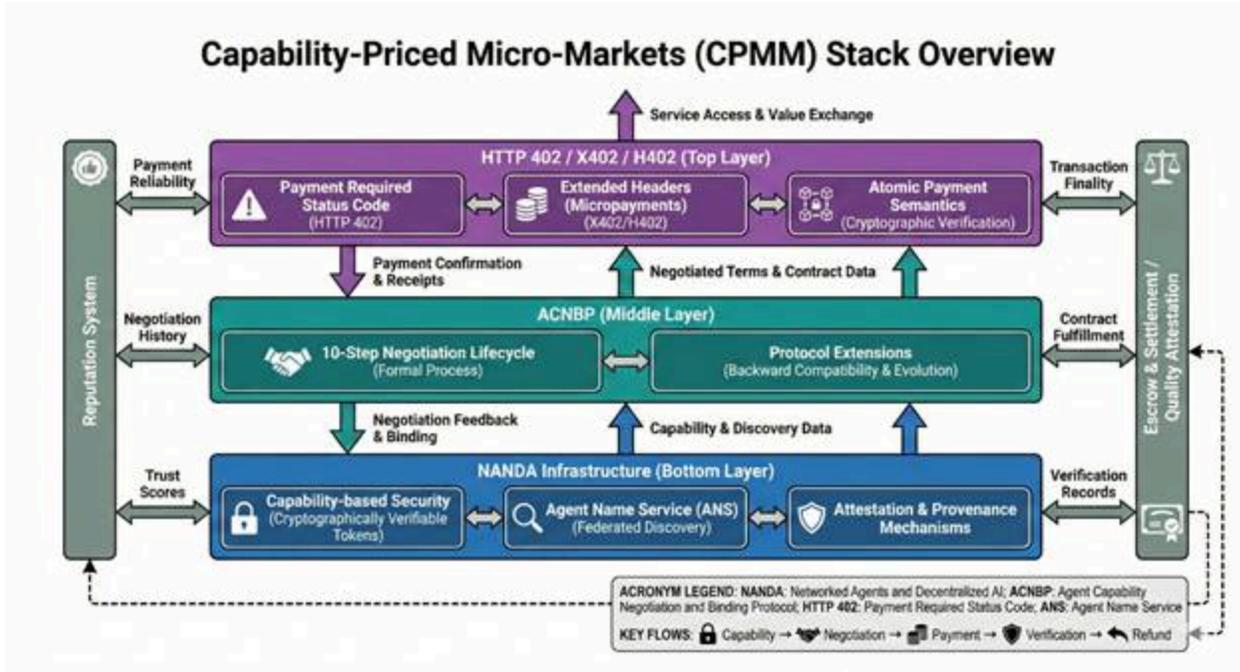

Figure 1 High Level CPMM stack overview

## 1.3 Theoretical Contributions

Beyond its practical contributions, this paper makes several theoretical advances in the economics of autonomous systems. We formalize the agent interaction problem as a repeated bilateral game with incomplete information, where agents must learn about each other's capabilities and costs through a sequence of interactions. This formalization enables us to prove convergence properties for the price discovery mechanism and to characterize the conditions under which efficient outcomes emerge.

Our main theoretical result demonstrates that under reasonable assumptions about agent rationality and information structure, the CPMM mechanism converges to a constrained Radner equilibrium—a generalization of competitive equilibrium that accounts for the information and capability constraints inherent in agent systems. This convergence occurs in $O(\log 1/\varepsilon)$ rounds for any desired precision $\varepsilon$, providing both theoretical guarantees and practical guidance for system design.

## 1.4 Paper Organization

The remainder of this paper is organized as follows. Section 2 provides detailed background on the three foundational technologies and establishes the threat model and trust assumptions. Section 3 develops the formal economic model, including agent preference structures, the repeated bilateral game formulation, and convergence analysis. Section 4 describes the protocol integration, showing how CPMM maps onto the existing ACNBP framework and detailing the economic payload structures. Section 5 analyzes the privacy economics, developing the theory of privacy elasticity and its implications for system design. Section 6 surveys related work and positions CPMM within the broader landscape of agent economics and distributed systems. Section 7 concludes with a discussion of implications and future research directions.

Throughout the paper, we maintain a focus on both theoretical rigor and practical implementability. Each theoretical result is accompanied by discussion of its implementation implications, and each protocol specification includes consideration of real-world deployment constraints. This dual focus reflects our belief that the future of autonomous agent systems requires both sound theoretical foundations and pragmatic engineering solutions.

## 2. Preliminaries

This section establishes the foundational technologies, formal models, and security assumptions that underpin the CPMM framework. We begin with a detailed exposition of the three core technologies—MIT NANDA, HTTP 402/X402/H402, and ACNBP—before developing the formal system model and threat assumptions.

### 2.1 MIT NANDA: Capability-Based Agent Infrastructure

MIT's NANDA(Networked Agents and Decentralized AI) represents a fundamental reimagining of agent infrastructure, moving beyond traditional identity-based security models to a capability-centric approach that aligns naturally with the functional nature of agent interactions [4]. The NANDAarchitecture provides four critical components that form the foundation for economic coordination: discovery mechanisms, authentication protocols, traceability systems, and a federated registry infrastructure.

#### 2.1.1 Capability-Based Security Model

The core innovation of NANDAlies in its adoption of capability-based security, a paradigm where access rights are represented as unforgeable tokens that combine both the authority and the means to access resources [7]. In the context of agent systems, a capability represents not just permission to invoke a service, but cryptographic proof of the agent's ability to provide that service.

Formally, a NANDAcapability is a tuple $C = \langle id, spec, proof, constraints \rangle$ where:

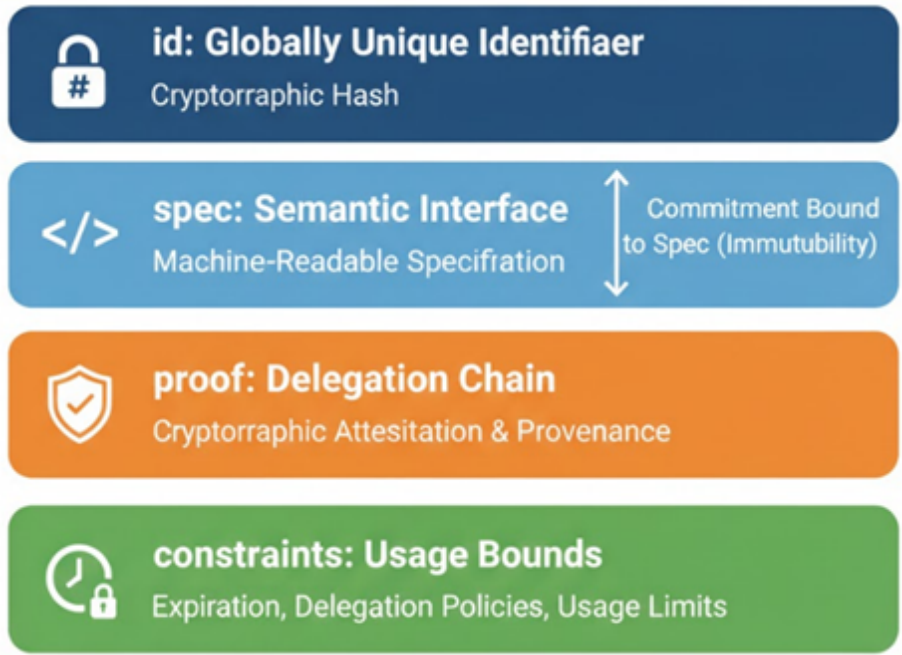

The capability specification spec follows a structured format that enables both human understanding and automated reasoning. For a natural language processing capability, the specification might include supported languages, accuracy guarantees, latency bounds, and input/output formats. Critically, these specifications are not mere documentation but form part of the cryptographic commitment, ensuring that agents cannot advertise capabilities they do not possess.

The proof component provides the cryptographic foundation for trust in the capability system. Capabilities are issued by trusted authorities (which may themselves be agents) and form chains of delegation that can be verified without contacting the issuing authority. This delegation model enables fine-grained capability distribution while maintaining security and auditability.

### 2.1.2 Agent Name Service (ANS)

The Agent Name Service provides a federated registry system that enables efficient agent discovery while preserving privacy and supporting the decentralized nature of agent ecosystems. Unlike traditional DNS, which maps names to network addresses, ANS maps agent identifiers to capability advertisements and service endpoints.

The ANS architecture consists of multiple registry nodes that maintain eventually consistent views of the agent namespace. Each registry node is responsible for a portion of the namespace, determined by a consistent hashing scheme that ensures load balancing and fault tolerance. Agents register with their local registry node, which propagates capability advertisements to other nodes based on demand and replication policies.

Privacy is preserved through selective capability disclosure, where agents can choose to advertise only a subset of their capabilities or to provide different levels of detail to different requesters. This selective disclosure is implemented through cryptographic commitment schemes that allow agents to prove possession of capabilities without revealing their full specifications.

The discovery process operates through a multi-stage protocol. First, a requesting agent queries the ANS with a capability specification describing its requirements. The ANS returns a list of candidate agents along with their advertised capabilities and service endpoints. The requesting agent can then initiate direct negotiation with promising candidates using the ACNBP protocol. The discovery process within the ANS must inherently support identity governance to mitigate the risks of silent cloning and behavioral drift. By operationalizing the Digital Identity Rights Framework (DIRF), the ANS can enforce 63 specific controls across nine domains, including consent verification and royalty triggers, to protect the biometric and personality-based likeness of agents [10]. This integration ensures that capability-based discovery is coupled with a robust model for identity traceability and monetization, which is essential as agents begin to maintain simulated portrayals of user identities over time.

### 2.1.3 Attestation and Verification Mechanisms

NANDA's attestation system provides cryptographic guarantees about agent capabilities and behavior. Attestations are generated by trusted execution environments (TEEs) or other secure hardware, ensuring that they cannot be forged or manipulated by malicious agents.

The attestation process involves several components:

**Capability Attestation**: When an agent claims to possess a capability, it must provide cryptographic proof that it has been granted that capability by an authorized issuer. This proof takes the form of a capability certificate that includes the capability specification, delegation chain, and digital signatures from all parties in the chain.

**Behavioral Attestation**: Beyond static capability possession, NANDAsupports attestation of dynamic behavior through execution logs and performance metrics. These attestations are generated by monitoring systems that observe agent interactions and provide cryptographic proof of service quality, response times, and adherence to service level agreements.

**Provenance Attestation**: The system maintains cryptographic records of capability provenance, enabling agents to verify not just what capabilities another agent possesses, but where those capabilities originated and how they were obtained. This provenance information is crucial for establishing trust in capability claims and detecting potential security violations.

### 2.1.4 Integration with Economic Mechanisms

The NANDAinfrastructure provides several features that are essential for economic coordination. The capability-based security model naturally aligns with economic valuation, as capabilities represent concrete, verifiable assets that can be priced and traded. The attestation mechanisms provide the trust foundation necessary for financial transactions, while the federated registry enables efficient market discovery.

Most importantly for CPMM, NANDA's delegation model enables sophisticated economic relationships. Capabilities can be delegated with usage constraints that include payment requirements, enabling agents to monetize their capabilities directly through the security infrastructure. This tight integration between security and economics reduces the complexity of implementing economic protocols and provides stronger guarantees about transaction integrity.

## 2.2 HTTP 402 and Micropayment Evolution

The HTTP 402 "Payment Required" status code represents one of the web's most prescient but underutilized features. Defined in the original HTTP/1.0 specification but marked as "reserved for future use," HTTP 402 was intended to support digital cash systems that were technically infeasible at the time of its creation [8]. Recent advances in cryptographic payment systems have finally made the original vision of frictionless web payments technically and economically viable.

### 2.2.1 HTTP 402 Foundation

The HTTP 402 status code provides a standardized mechanism for servers to indicate that payment is required to access a resource. Unlike other HTTP error codes that indicate technical problems, 402 represents a business logic condition that can be resolved through payment. The original specification was intentionally minimal, providing only the status code itself without defining payment mechanisms or protocols.

This minimalism has proven to be both a strength and a limitation. The lack of prescribed payment mechanisms has allowed for innovation in payment protocols while maintaining compatibility with existing HTTP infrastructure. However, it has also led to fragmentation, with different implementations using incompatible payment schemes.

The revival of interest in HTTP 402 has been driven by several technological developments: the emergence of cryptocurrencies and digital payment systems that can handle micro-transactions economically, advances in cryptographic protocols that enable secure, low-latency payment verification, and the growth of API economies where fine-grained usage-based pricing is economically important.

### 2.2.2 X402/H402 Protocol Extensions

The X402 and H402 protocols represent modern extensions to HTTP 402 that address its original limitations while maintaining backward compatibility [5]. These protocols introduce structured payment metadata through HTTP headers, enabling rich payment semantics without requiring modifications to existing HTTP infrastructure.

The X402 protocol introduces several key headers:

**X402-Payment-Required**: Indicates the payment amount and currency, along with payment method preferences and timeout constraints. The header format supports multiple currencies and payment methods, enabling clients to choose their preferred payment mechanism.

**X402-Payment-Address**: Specifies the payment destination, which may be a cryptocurrency address, payment processor endpoint, or other payment identifier. The address format is extensible, supporting both current and future payment systems.

**X402-Payment-Metadata**: Provides additional payment context, including invoice identifiers, service level agreements, and refund policies. This metadata enables sophisticated payment relationships that go beyond simple pay-per-request models.

The H402 protocol builds on X402 by adding cryptographic payment verification directly in HTTP headers. Key innovations include:

**Ephemeral Key Management**: Each payment transaction uses ephemeral Ed25519 key pairs, ensuring that payment credentials cannot be reused or stolen. The public key is transmitted in the H402-Payment-Key header, while the private key is used to sign the payment commitment.

**Compact Signatures**: Payment commitments are signed using Schnorr signatures, which provide strong security guarantees while requiring only 64 bytes of header space. This compact representation minimizes protocol overhead while maintaining cryptographic security.

**Atomic Payment Semantics**: The H402 protocol ensures that payments are atomic with respect to service delivery. Payments are committed only when services are successfully delivered, and automatic refunds are triggered by service failures.

### 2.2.3 Economic Properties of Micropayments

The economic viability of micropayments depends critically on the relationship between transaction value and transaction cost. Traditional payment systems impose fixed costs (typically $0.20-$0.30 per transaction) that make small-value transactions economically infeasible. The X402/H402 protocols address this challenge through several mechanisms:

**Cryptographic Settlement**: By using cryptographic proofs rather than traditional payment processing, transaction costs are reduced to the computational cost of signature verification, typically measured in microseconds rather than dollars.

**Batching and Aggregation**: The protocols support payment batching, where multiple small transactions can be aggregated into larger settlements, amortizing fixed costs across multiple transactions.

**Conditional Payments**: The atomic payment semantics enable conditional payments that are only settled upon successful service delivery, reducing the risk of payment for undelivered services.

These properties enable economically viable transactions at the scale required for agent interactions, where individual service invocations might be worth fractions of a cent but occur at high frequency.

## 2.3 Agent Capability Negotiation and Binding Protocol (ACNBP)

ACNBP provides the structured negotiation framework that enables agents to discover, evaluate, and commit to service relationships [6]. The protocol's innovation lies in its formal approach to negotiation, providing cryptographic guarantees about the integrity and enforceability of negotiation outcomes.

### 2.3.1 Protocol Architecture

ACNBP is structured as a 10-step protocol that guides agents through the complete lifecycle of service negotiation and delivery:

1. **Discover**: Agents use the ANS to identify potential service providers based on capability requirements.

2. **Pre-screen**: Initial filtering of candidates based on basic compatibility criteria, availability, and preliminary cost estimates.

3. **Negotiate-Request**: Formal negotiation initiation, where the requesting agent submits detailed requirements and constraints.

4. **Negotiate-Response**: Provider agents respond with capability confirmations, pricing, and service level commitments.

5. **Bind**: Mutual commitment to the negotiated terms, creating enforceable service agreements.

6. **Commit**: Cryptographic commitment to the service delivery, including payment escrow and performance bonds.

7. **Execute**: Actual service delivery, with real-time monitoring and attestation generation.

8. **Verify**: Post-delivery verification of service quality and compliance with negotiated terms.

9. **Release**: Settlement of payments and release of any held resources or bonds.

10. **Audit**: Generation of audit trails and reputation updates based on transaction outcomes.

Each step in the protocol is cryptographically secured, with digital signatures ensuring message integrity and non-repudiation. The protocol's state machine design ensures that agents cannot skip steps or manipulate the negotiation process.

See Figure 2 (Setup and Configure) and Figure 3 for simulated transaction log based on Figure 2:

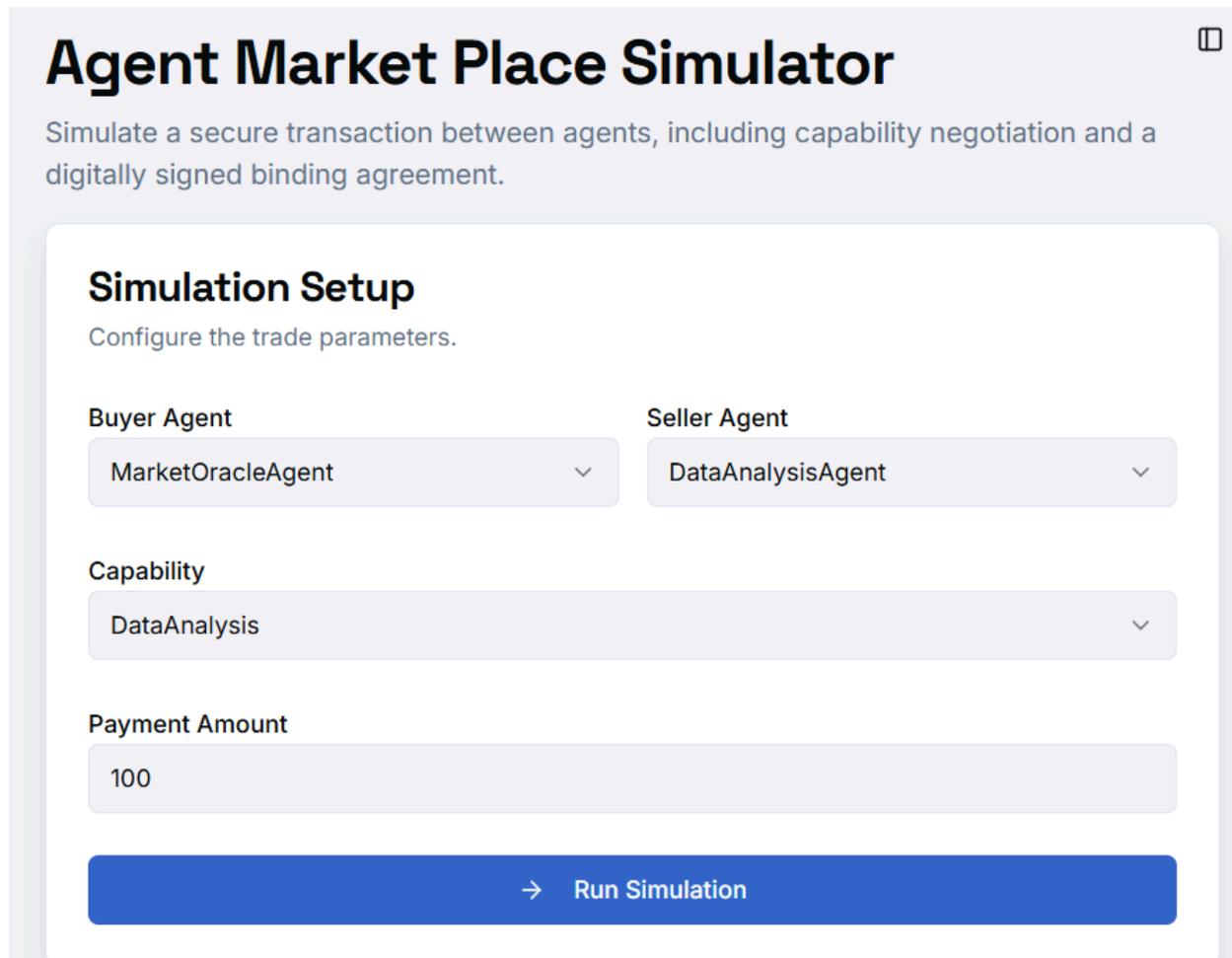

Figure 2: Agent Market Place Simulator

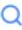

**Figure 3: Simulation Transaction Logs**

### 2.3.2 Capability Advertisement and Discovery

The capability advertisement mechanism in ACNBP extends the basic ANS discovery with rich semantic descriptions and dynamic pricing information. Capability advertisements include:

**Functional Specifications**: Detailed descriptions of what the capability does, including input/output formats, supported operations, and behavioral constraints.

**Non-Functional Properties**: Performance characteristics such as latency bounds, throughput limits, availability guarantees, and accuracy metrics.

**Economic Parameters**: Pricing models, payment preferences, service level agreements, and penalty structures for non-compliance.

**Dynamic State**: Real-time information about current load, availability, and performance metrics.

The discovery process uses semantic matching algorithms to identify compatible capabilities, taking into account both functional requirements and economic constraints. This matching process is designed to be efficient even in large-scale agent networks, using indexing and caching strategies to minimize discovery latency.

### 2.3.3 Negotiation Semantics

The negotiation phase of ACNBP supports sophisticated multi-attribute negotiation, where agents can negotiate not just on price but on service quality, delivery timing, and other parameters. The negotiation protocol supports several interaction patterns:

**Bilateral Negotiation**: Direct negotiation between a service requester and a single provider, suitable for specialized services or when specific provider characteristics are required.

**Multi-party Negotiation**: Negotiation involving multiple providers, enabling competitive bidding and service composition scenarios.

**Iterative Refinement**: Multi-round negotiation where agents can refine their requirements and offers based on market feedback.

The protocol ensures fairness through cryptographic commitment schemes that prevent agents from changing their offers after seeing competitors' bids. This commitment mechanism is essential for maintaining market integrity and preventing strategic manipulation.

### 2.3.4 Binding and Enforcement

The binding phase creates enforceable agreements between agents, using cryptographic contracts that specify service delivery requirements, payment terms, and penalty structures. These contracts are designed to be automatically enforceable without requiring human intervention or external arbitration.

The enforcement mechanism relies on several components:

**Cryptographic Escrow**: Payments are held in cryptographic escrow during service delivery, ensuring that providers are paid for successful delivery while protecting requesters from non-delivery.

**Attestation-Based Verification**: Service delivery is verified through cryptographic attestations generated by trusted execution environments, providing objective evidence of compliance with service agreements.

**Automated Dispute Resolution**: Disputes are resolved automatically based on cryptographic evidence, eliminating the need for human arbitrators and reducing resolution time.

**Reputation Integration**: Transaction outcomes are automatically recorded in reputation systems, providing long-term incentives for honest behavior.

## 2.4 System Model and Assumptions

Having established the foundational technologies, we now develop the formal system model that underlies the CPMM framework. This model abstracts the essential features of agent interactions while remaining grounded in the practical constraints of real systems.

### 2.4.1 Agent Model

We model the agent ecosystem as a dynamic network $G = (A, E, T)$ where $A$ is the set of agents, $E$ represents potential interaction relationships, and $T$ captures the temporal evolution of the network. Each agent $a \in A$ is characterized by:

**Capability Set**: $C_a \subseteq C$, where $C$ is the universe of possible capabilities. Each capability $c \in C_a$ has an associated cost function $cost_a(c, q, s)$ that depends on service quality $q$ and agent state $s$.

**Preference Structure**: A utility function $u_a(q, p, s)$ that captures the agent's valuation of service quality $q$ at price $p$ given state $s$. We assume that utility functions are continuous, monotonic in quality, and decreasing in price.

**Information Structure**: Each agent has private information about its costs and capabilities, and partial information about other agents obtained through past interactions and market observations.

**Strategic Behavior**: Agents are assumed to be rational utility maximizers with bounded computational resources. They may use learning algorithms to adapt their strategies based on market experience.

### 2.4.2 Interaction Model

Agent interactions follow the ACNBP protocol structure, with each interaction consisting of a discovery phase, negotiation phase, and execution phase. We model interactions as discrete events that occur in continuous time, with agents able to engage in multiple simultaneous interactions.

The interaction model includes several important features:

**Asynchronous Communication**: Agents communicate through message passing with variable delays, reflecting the realities of distributed systems.

**Partial Observability**: Agents can only observe their own interactions and limited market information, creating information asymmetries that affect pricing and negotiation strategies.

**Dynamic Capabilities**: Agent capabilities can change over time due to learning, resource constraints, or external factors.

**Network Effects**: The value of capabilities may depend on network structure and the availability of complementary services.

### 2.4.3 Economic Environment

The economic environment is characterized by several key parameters:

**Market Structure**: The degree of competition in different capability markets, ranging from monopolistic (single provider) to perfectly competitive (many identical providers).

**Information Asymmetry**: The extent to which agents have private information about their costs, capabilities, and valuations.

**Transaction Costs**: The costs associated with discovery, negotiation, and settlement, which affect the economic viability of different interaction patterns.

**Network Externalities**: The degree to which the value of services depends on the number and types of other agents in the network.

To secure the high-velocity economic environment enabled by H402, the framework aligns with a unified Zero-Trust architecture designed specifically for the Agentic Web [11]. This architecture utilizes Decentralized Identifiers (DIDs) and Verifiable Credentials (VCs) to anchor agent identities, ensuring that every micropayment interaction is continuously authenticated and authorized. By incorporating innovations such as Trust-Adaptive Runtime Environments (TARE) and Causal Chain Auditing, the system provides provable security guarantees against logic-layer threats that might otherwise manipulate economic payload structures or compromise payment atomicity.

### 2.5 Threat Model and Security Assumptions

The security of the CPMM framework depends on several assumptions about the threat environment and the capabilities of adversarial agents. We adopt the MAESTRO threat modeling framework [9] and extend it to address economic threats specific to agent marketplaces.

### 2.5.1 Adversarial Capabilities

We consider adversaries with the following capabilities:

**Network Control**: Adversaries may control network infrastructure, enabling them to intercept, delay, or modify messages between agents. However, we assume that cryptographic primitives remain secure and that adversaries cannot break digital signatures or decrypt properly encrypted communications.

**Agent Compromise**: Adversaries may compromise individual agents, gaining access to their private keys, internal state, and communication channels. However, we assume that the compromise of one agent does not automatically lead to the compromise of others.

**Economic Manipulation**: Adversaries may attempt to manipulate market prices through various strategies, including Sybil attacks (creating multiple fake identities), wash trading (artificial transaction volume), and predatory pricing.

**Capability Misrepresentation**: Adversaries may attempt to advertise capabilities they do not possess or to provide services that do not meet advertised specifications.

The threat model must be extended to include Logic-layer Prompt Control Injection (LPCI), a fundamentally different attack vector that exploits the planning and orchestration mechanisms of autonomous agents [12]. LPCI represents a paradigm shift from surface-level manipulation to long-term system compromise, where dormant, encoded payloads are embedded within persistent memory or vector databases. These attacks bypass conventional input filters and can be triggered across sessions by keywords or tool outputs, effectively turning the agent into an unwitting accomplice in unauthorized transactions or data exfiltration. Mitigating LPCI requires moving beyond static filtering toward runtime enforcement of logic-layer integrity and cryptographic tool attestation.

### 2.5.2 Trust Anchors and Assumptions

The security of CPMM relies on several trust anchors:

**Cryptographic Primitives**: We assume the security of standard cryptographic primitives, including digital signatures, hash functions, and encryption schemes. Specifically, we assume that Ed25519 signatures are unforgeable and that SHA-256 is collision-resistant.

**Trusted Execution Environments**: The attestation mechanisms rely on trusted execution environments (TEEs) such as Intel SGX or ARM TrustZone. We assume that these environments provide integrity and confidentiality guarantees for code execution and that attestations generated by TEEs are trustworthy.

**Capability Authorities**: The capability-based security model relies on trusted capability authorities that issue and revoke capabilities. We assume that these authorities are honest and that their private keys are secure.

**Network Assumptions**: We assume that the network provides eventual message delivery, though messages may be delayed, reordered, or lost. We do not assume synchrony or reliable broadcast.

### 2.5.3 Economic Threat Analysis

Beyond traditional security threats, CPMM must address several economic attack vectors:

**Sybil Attacks**: Adversaries may create multiple fake agent identities to manipulate market prices or to gain unfair advantages in negotiations. The capability-based security model provides some protection against Sybil attacks by requiring cryptographic proof of capability possession, but additional economic mechanisms are needed to fully address this threat.

**Price Manipulation**: Adversaries may attempt to manipulate prices through coordinated bidding strategies, artificial scarcity creation, or market cornering. The decentralized nature of the CPMM marketplace provides some resistance to price manipulation, but monitoring and detection mechanisms are still necessary.

**Service Quality Attacks**: Adversaries may provide low-quality services while claiming to meet advertised specifications, potentially damaging the reputation of the marketplace and reducing overall efficiency. The attestation-based verification mechanisms provide protection against this threat, but economic incentives must be aligned to encourage honest service provision.

**Payment Attacks**: Adversaries may attempt to receive services without payment or to double-spend payment tokens. The cryptographic payment mechanisms and escrow systems provide protection against these attacks, but careful protocol design is required to ensure atomicity of payment and service delivery.

This comprehensive threat model guides the design of CPMM's security mechanisms and informs the economic analysis that follows. By explicitly considering both technical and economic threats, we can design a system that is robust against a wide range of adversarial behaviors while maintaining efficiency and usability for honest participants.

## 3. Economic Model and Theoretical Foundations

This section develops the formal economic theory underlying the CPMM framework. We begin by establishing agent preference structures and cost models, then formalize the market interaction as a repeated bilateral game with incomplete information. Our main theoretical contributions include convergence analysis for the price discovery mechanism and characterization of equilibrium properties under various market conditions.

## 3.1 Agent Preference and Cost Models

The foundation of any economic analysis lies in the specification of agent preferences and cost structures. In the context of autonomous agents, these preferences must capture both the functional requirements of service consumption and the resource constraints of service provision.

### 3.1.1 Service Quality and Capability Specifications

We model service quality as a multi-dimensional vector $q \in Q \subseteq \mathbb{R}^d$, where each dimension represents a measurable aspect of service performance. For a natural language processing service, quality dimensions might include accuracy (measured by BLEU score), latency (response time in milliseconds), and throughput (requests per second). The quality space Q is assumed to be a compact, convex subset of $\mathbb{R}^d$, ensuring that quality levels are bounded and that interpolation between quality levels is meaningful.

Each agent i possesses a capability set $C_i$, where each capability $c \in C_i$ is characterized by a feasible quality region $Q_c \subseteq Q$. The feasible quality region represents the range of quality levels that the agent can deliver for that capability, potentially depending on current load, resource availability, and other dynamic factors.

Formally, we define a capability as a tuple c = ⟨spec, $Q_c$, cost⟩ where:

- spec is a semantic specification of the capability's functionality
- $Q_c \subseteq Q$ is the feasible quality region
- cost: $Q_c \times S \to \mathbb{R}_+$ is a cost function mapping quality levels and agent states to non-negative costs

The agent state $s \in S$ captures dynamic factors that affect service provision, including current computational load, available memory, energy levels, and network connectivity. The state space S is assumed to be a measurable space with an associated probability measure reflecting the stochastic evolution of agent states over time.

### 3.1.2 Buyer Preference Structure

A buyer agent's preferences are represented by a utility function u: $Q \times \mathbb{R}_+ \times S \to \mathbb{R}$ that maps service quality, price, and buyer state to utility levels. We impose several regularity conditions on utility functions to ensure well-behaved economic analysis:

**Monotonicity**: For any fixed price p and state s, utility is non-decreasing in quality: $q' \geq q$ implies u(q', p, s) ≥ u(q, p, s).

**Price Sensitivity**: For any fixed quality q and state s, utility is non-increasing in price: $p' \geq p$ implies u(q, p', s) ≤ u(q, p, s).

**Continuity**: The utility function is continuous in all arguments, ensuring that small changes in quality or price lead to small changes in utility.

**Concavity**: Utility is concave in quality, reflecting diminishing marginal returns to quality improvements.

A common specification that satisfies these conditions is the quasi-linear utility function:

$$u(q, p, s) = v(q, s) - p$$

where $v(q, s)$ is a valuation function that captures the buyer's willingness to pay for quality level q in state s. The quasi-linear form simplifies analysis while capturing the essential trade-off between quality and price.

For multi-dimensional quality, we often use separable valuation functions:

$$v(q, s) = \sum_{j=1}^{d} w_j(s) \cdot v_j(q_j)$$

where $w_j(s)$ represents the state-dependent weight placed on quality dimension j, and $v_j$ is a univariate valuation function for that dimension.

### 3.1.3 Seller Cost Structure

Seller agents face costs that depend on both the quality level delivered and their current state. We model the cost function as:

$$cost(q, s) = fc(s) + vc(q, s)$$

where $fc(s)$ represents fixed costs that depend on agent state but not on service quality, and $vc(q, s)$ represents variable costs that increase with quality requirements.

The variable cost function typically exhibits increasing marginal costs, reflecting the fact that achieving higher quality levels requires disproportionately more resources. A common specification is:

$$vc(q, s) = \sum_{j=1}^{d} \alpha_j(s) \cdot q_j^{\beta_j}$$

where $\alpha_j(s)$ represents the state-dependent cost coefficient for quality dimension j, and $\beta_j > 1$ captures increasing marginal costs.

The fixed cost component $fc(s)$ captures setup costs, opportunity costs of capacity allocation, and other costs that are incurred regardless of the specific quality level delivered. This component is crucial for understanding the economics of agent participation, as agents will only participate in markets where expected revenues exceed fixed costs.

### 3.1.4 Information Structure and Beliefs

A critical aspect of the economic model is the specification of information structure and agent beliefs. We assume that each agent has private information about its own costs and valuations, while having only partial information about other agents.

Specifically, each buyer agent knows its own valuation function $v(q, s)$ but observes only noisy signals about seller costs. Similarly, each seller agent knows its own cost function $cost(q, s)$ but observes only market prices and rejection rates to infer buyer valuations.

We model beliefs using Bayesian updating, where agents maintain probability distributions over unknown parameters and update these distributions based on observed market outcomes. For a buyer agent, beliefs about seller costs are represented by a probability distribution $F(c \mid \theta)$ where $\theta$ represents the buyer's belief parameters, updated according to Bayes' rule based on observed prices and service quality.

$$F_{t+1}(c|\theta) = F_t(c|\theta) \times L(observation|c) / \int F_t(c'|\theta) \times L(observation|c') \, dc'$$

The information structure creates several important economic phenomena:

**Adverse Selection**: Agents with private information about their types may have incentives to misrepresent their characteristics, leading to market inefficiencies.

**Learning Dynamics**: Agents learn about market conditions through repeated interactions, leading to evolving pricing and negotiation strategies.

**Signaling**: Agents may use their pricing and quality choices to signal their private information to other market participants.

## 3.2 Repeated Bilateral Game Formulation

The core of the CPMM mechanism is a repeated bilateral game where buyer and seller agents interact over time to discover mutually beneficial trading opportunities. This section formalizes the game structure and analyzes its equilibrium properties.

### 3.2.1 Stage Game Structure

In each period t, a buyer agent is matched with a seller agent according to a matching process that depends on the ANS discovery mechanism and agent preferences. The stage game proceeds as follows:

1. **Capability Advertisement**: The seller announces a capability specification and quality region $Q_c$.

2. **Quality Request**: The buyer submits a quality requirement $q \in Q_c$ and a maximum willingness to pay $\bar{p}$.

3. **Price Quote**: The seller responds with either a price quote p or a rejection (HTTP 402 response).

4. **Accept/Reject Decision**: If a price quote is received, the buyer decides whether to accept or reject the offer.

5. **Service Delivery**: If the offer is accepted, the seller delivers the service and receives payment.

The stage game payoffs depend on the outcome:

- If no trade occurs, both agents receive payoff 0.
- If trade occurs at price p with quality q, the buyer receives utility u(q, p, s_b) and the seller receives profit p - cost(q, s_s).

### 3.2.2 Matching Process and Market Structure

The matching process determines which buyer-seller pairs interact in each period. We consider several matching mechanisms:

**Random Matching**: Buyers and sellers are matched uniformly at random from the pool of active agents. This provides a baseline for analysis but may not reflect realistic discovery patterns.

**Capability-Based Matching**: The ANS discovery mechanism matches buyers with sellers based on capability compatibility, potentially with some randomness to ensure market exploration.

**Preference-Based Matching**: Agents can express preferences over potential trading partners based on past experience, reputation, or other factors.

The matching probability between buyer i and seller j in period t is denoted $\alpha_{ij}(t)$, which may depend on agent characteristics, past interaction history, and market conditions.

### 3.2.3 Learning and Adaptation

Agents adapt their strategies over time based on observed market outcomes. We model this adaptation using a learning framework where agents maintain beliefs about market conditions and update these beliefs based on experience.

For buyer agents, the learning process involves updating beliefs about seller cost distributions based on observed prices and rejection rates. Let $F_t(c)$ denote the buyer's belief about the seller cost distribution at time t. After observing a price quote p or rejection, the buyer updates beliefs according to:

$$F_{t+1}(c) = F_t(c \mid \text{observed outcome})$$

using Bayes' rule. This learning process affects future bidding strategies, as buyers adjust their willingness to pay based on their evolving understanding of market conditions.

Seller agents similarly learn about buyer valuations by observing acceptance and rejection decisions. This learning affects pricing strategies, as sellers adjust their price quotes based on their understanding of buyer demand.

The stability of the Radner equilibrium is contingent on the cognitive health of participating agents. Internal failures such as memory starvation, planner recursion, and context flooding can lead to 'Cognitive Degradation,' a novel vulnerability class where an agent's reasoning capability silently drifts or collapses. To maintain the integrity of the learning and adaptation process, CPMM integrates the Qorvex Security AI Framework (QSAF). QSAF provides seven runtime controls, such as starvation detection and fallback logic rerouting, that monitor agent subsystems in real time. This ensures that agents remain resilient against internal logic collapse, which is essential for consistent price discovery and market stability in multi-agent environments [13].

### 3.2.4 Strategic Behavior and Equilibrium Concepts

We analyze agent behavior using the concept of Bayesian Nash equilibrium, where each agent's strategy is optimal given their beliefs about other agents' strategies and types. In the repeated game setting, we focus on stationary equilibria where strategies depend only on current beliefs and market conditions, not on calendar time.

A strategy for a buyer agent specifies:

- A quality requirement function $q(s_b, F_t)$ mapping buyer state and beliefs to quality requests
- A bidding function $\bar{p}(s_b, F_t)$ mapping buyer state and beliefs to maximum willingness to pay
- An acceptance rule $\alpha(p, q, s_b, F_t)$ determining whether to accept price quotes

A strategy for a seller agent specifies:

- A capability advertisement policy determining which capabilities to advertise and how to describe them
- A pricing function $p(q, s_s, G_t)$ mapping quality requests, seller state, and beliefs to price quotes
- A rejection rule determining when to reject requests (send HTTP 402 responses)

The equilibrium concept requires that each agent's strategy is optimal given the strategies of other agents and the matching process. This leads to a fixed-point condition where beliefs are consistent with observed behavior and strategies are optimal given beliefs.

## 3.3 Convergence Analysis and Equilibrium Properties

This section presents our main theoretical results on the convergence properties of the CPMM mechanism and characterizes the equilibrium outcomes under various market conditions.

### 3.3.1 Convergence to Radner Equilibrium

Our central theoretical result demonstrates that the CPMM mechanism converges to a constrained Radner equilibrium under reasonable assumptions about agent behavior and market structure.

**Definition 3.1 (Constrained Radner Equilibrium)**: A constrained Radner equilibrium is a tuple $(p^*, q^*, \mu^*)$ where $p^*$ is a price function, $q^*$ is a quality allocation, and $\mu^*$ is a belief system such that:

1. Given beliefs $\mu^*$, the price and quality choices $(p^*, q^*)$ constitute a Nash equilibrium
2. Beliefs $\mu^*$ are consistent with observed behavior under $(p^*, q^*)$
3. The allocation satisfies capability constraints: $q^* \in \cap_i Q_i$ for all active agents i

The constrained Radner equilibrium generalizes the standard competitive equilibrium to account for the information constraints and capability limitations inherent in agent systems.

**Theorem 3.1 (Convergence to Equilibrium)**: Under the following assumptions:

1. Agent utility and cost functions are L-Lipschitz continuous
2. The matching probability satisfies $\alpha_t \geq c/t$ for some constant c > 0
3. Agents use Bayesian learning with bounded memory of length k
4. The capability space is compact and convex

The sequence of market prices $\{p_t\}$ converges to the equilibrium price $p^*$ at rate $O(\log(1/\varepsilon)/t)$, where ε is the desired precision.

**Proof Sketch**: The proof proceeds in several steps:

1. **Contraction Mapping**: We show that the belief updating process defines a contraction mapping on the space of probability distributions over agent types.

2. **Stochastic Approximation**: The learning dynamics can be analyzed as a stochastic approximation algorithm, where the belief updates correspond to gradient steps toward the equilibrium.

3. **Martingale Convergence**: Using martingale convergence theorems, we establish that the belief sequence converges almost surely to the equilibrium beliefs.

4. **Price Convergence**: Given belief convergence, the optimality conditions for agent strategies imply convergence of prices to equilibrium levels.

The logarithmic convergence rate reflects the information-theoretic complexity of learning in markets with private information. The dependence on the matching probability $\alpha_t$ captures the role of market thickness in facilitating price discovery.

### 3.3.2 Efficiency Properties

The efficiency of the CPMM mechanism depends on how closely the equilibrium allocation approximates the first-best outcome that would be achieved with complete information.

**Definition 3.2 (Allocative Efficiency)**: The allocative efficiency of an outcome is measured by the ratio of achieved social welfare to first-best social welfare:

$$\eta = (\Sigma_i u_i(q_i, p_i, s_i) + \Sigma_j (p_j - cost_j(q_j, s_j))) / W^*$$

where $W^*$ is the first-best social welfare.

**Theorem 3.2 (Efficiency Bounds)**: Under the assumptions of Theorem 3.1, the allocative efficiency of the CPMM mechanism satisfies:

$$\eta \geq 1 - O(\sqrt{\log n / n})$$

where $n$ is the number of agents in the market.

This result shows that the efficiency loss due to information constraints decreases as the market size increases, reflecting the benefits of market thickness for information aggregation.

### 3.3.3 Robustness to Strategic Manipulation

A critical concern in any market mechanism is robustness to strategic manipulation by participants. We analyze several forms of manipulation and show that the CPMM mechanism provides strong resistance to common attack strategies.

**Theorem 3.3 (Sybil Resistance)**: Under the capability-based security model, the benefit to an agent from creating k Sybil identities is bounded by:

$$B(k) \leq O(\log k / \sqrt{k})$$

This sublinear bound implies that the marginal benefit of additional Sybil identities decreases rapidly, making large-scale Sybil attacks economically unattractive.

**Proof Sketch**: The capability-based security model requires each agent identity to possess verifiable capabilities. Since capabilities are scarce resources that cannot be arbitrarily replicated, the cost of creating Sybil identities grows linearly with the number of identities. Meanwhile, the benefit from market manipulation grows sublinearly due to the decentralized nature of the matching process.

## 3.4 Combinatorial Auctions for Workflow Composition

Many agent interactions involve complex workflows that require coordination among multiple service providers. This section extends the bilateral trading model to handle combinatorial auctions for workflow composition.

### 3.4.1 Workflow Representation

A workflow is represented as a directed acyclic graph $W = (V, E)$ where vertices $V$ represent tasks and edges $E$ represent dependencies between tasks. Each task $v \in V$ is associated with a capability requirement $c_v$ and quality specification $q_v$.

The workflow completion requires finding an assignment of agents to tasks such that:

1. Each assigned agent possesses the required capability for its task
2. Quality requirements are satisfied for all tasks
3. Dependency constraints are respected
4. The total cost is minimized subject to budget constraints

### 3.4.2 Auction Mechanism Design

We design a combinatorial auction mechanism that allows agents to bid on subsets of tasks, enabling efficient allocation even when individual agents cannot complete entire workflows.

**Bid Structure**: Each bid is a tuple $b = \langle T, q, p \rangle$ where:

- $T \subseteq V$ is a subset of tasks
- $q = \{q_v\}_{v \in T}$ specifies quality levels for each task in $T$
- $p$ is the total price for completing all tasks in $T$ at the specified quality levels

**Winner Determination**: The winner determination problem is formulated as an integer linear program:

minimize $\sum_\beta p_\beta \cdot x_\beta$

subject to:

- $\sum_{\beta : v \in T_\beta} x_\beta = 1$ for all $v \in V$ (each task assigned exactly once)
- $x_\beta \in \{0, 1\}$ for all bids $b$ (binary assignment variables)

- Budget and quality constraints

**Payment Rule**: We use a modified VCG (Vickrey-Clarke-Groves) payment rule that ensures truthful bidding while respecting the budget constraints imposed by HTTP 402 Max-Price headers.

The payment to winning bidder i is:

$p_i$ = (social welfare without i) - (social welfare with i, excluding i's contribution)

subject to the constraint that total payments do not exceed the buyer's declared budget.

### 3.4.3 Incentive Properties

**Theorem 3.4 (Truthfulness)**: Under the modified VCG payment rule with budget constraints, truthful bidding is a dominant strategy for all agents.

**Theorem 3.5 (Individual Rationality)**: All winning agents receive non-negative utility, and all losing agents receive zero utility.

**Theorem 3.6 (Budget Feasibility)**: Total payments never exceed the buyer's declared budget, ensuring compatibility with HTTP 402 payment constraints.

These properties ensure that the combinatorial auction mechanism is compatible with the broader CPMM framework while providing strong incentive guarantees.

## 3.5 Dynamic Pricing and Learning

The static analysis above provides important insights into equilibrium behavior, but real agent systems operate in dynamic environments where costs, capabilities, and demand patterns change over time. This section analyzes dynamic pricing strategies and their convergence properties.

### 3.5.1 Multi-Armed Bandit Formulation

We model the dynamic pricing problem as a multi-armed bandit where each "arm" corresponds to a different pricing strategy. The agent's goal is to learn the optimal pricing strategy while balancing exploration (trying new prices) with exploitation (using known good prices).

Let $\pi = \{\pi_1, \pi_2, ..., \pi_k\}$ be a set of pricing policies, where each $\pi_j$ maps market conditions to price quotes. The agent observes rewards $r(\pi_j, t)$ when using policy $\pi_j$ at time t, where rewards depend on whether trades are completed and at what prices.

The agent's problem is to select a sequence of policies $\{\pi_t\}$ to maximize cumulative reward:

maximize $\Sigma_{t=1}^{T} E[r(\pi_t, t)]$

subject to learning constraints and the need to balance exploration and exploitation.

### 3.5.2 Upper Confidence Bound Algorithm

We propose an Upper Confidence Bound (UCB) algorithm adapted for the pricing context:

**Algorithm 3.1 (UCB Pricing)**:

1. Initialize: For each policy $\pi_j$, $set\ \hat{\mu}_j(0) = 0\ and\ n_j(0) = 0$
2. For t = 1, 2, ..., T:
   a. Select policy $\pi_t = argmax_j\ [\hat{\mu}_j(t-1) + \sqrt{(2\ log\ t\ /\ n_j(t-1))}]$
   b. Observe reward $r(\pi_t,\ t)$
   c. Update: $n_t(t) = n_t(t-1) + 1$, $\hat{\mu}_t(t) = (\hat{\mu}_t(t-1) \cdot n_t(t-1) + r(\pi_t,\ t))\ /\ n_t(t)$

**Theorem 3.7 (Regret Bound)**: The UCB pricing algorithm achieves regret bounded by:

R(T) ≤ O(√(K log T · T))

where K is the number of pricing policies and T is the time horizon.

This logarithmic dependence on T shows that the algorithm efficiently learns the optimal pricing strategy.

### 3.5.3 Contextual Bandits for State-Dependent Pricing

In practice, optimal pricing strategies depend on market context such as current demand, agent load, and competitive conditions. We extend the bandit formulation to handle contextual information.

Let x(t) ∈ X represent the context (market state) at time t. The agent maintains a policy function π: X → ℝ₊that maps contexts to prices. We use a linear model:

π(x) = θᵀx

where θ is a parameter vector learned over time.

**Algorithm 3.2 (LinUCB Pricing)**:

1. Initialize: A₀ = I, b₀ = 0
2. For t = 1, 2, ..., T:
   a. Observe context x(t)
   b. Compute $\hat{\theta}_t = A_{t-1}^{-1} b_{t-1}$
   c. Set price $p(t) = \hat{\theta}_t^T x(t) + \alpha\sqrt{(x(t)^T A_{t-1}^{-1} x(t))}$
   d. Observe reward r(t) e. Update: $A_t = A_{t-1} + x(t)x(t)^T$, $b_t = b_{t-1} + r(t)x(t)$

**Theorem 3.8 (Contextual Regret Bound)**: Under appropriate regularity conditions, the LinUCB pricing algorithm achieves regret:

$$R(T) \leq O(d\sqrt{T \log T})$$

where d is the dimension of the context space.

This result shows that contextual learning can significantly improve pricing performance when market conditions vary systematically.

The theoretical framework developed in this section provides the foundation for understanding the economic properties of the CPMM mechanism. The convergence results demonstrate that the system can achieve efficient outcomes despite information constraints, while the robustness analysis shows that the mechanism is resistant to common forms of strategic manipulation. The dynamic pricing analysis provides practical algorithms for agents to adapt to changing market conditions while maintaining theoretical performance guarantees.

# 4. Protocol Integration and System Architecture

This section details how the CPMM economic framework integrates with the existing ACNBP protocol infrastructure, specifying the concrete mechanisms for economic coordination within the established agent communication patterns. We describe the mapping of economic concepts onto protocol messages, the structure of economic payloads, and the implementation of payment and dispute resolution mechanisms.

## 4.1 CPMM-ACNBP Integration Architecture

The integration of CPMM with ACNBP leverages the existing 10-step protocol structure while extending it with economic semantics and payment capabilities. This approach ensures backward compatibility with existing ACNBP implementations while enabling rich economic interactions for agents that support the CPMM extensions.

### 4.1.1 Protocol Extension Mechanism

ACNBP's protocolExtension mechanism provides the foundation for integrating economic functionality without breaking existing implementations. The CPMM extension is identified by the extension identifier "org.cpmm.economic-coordination.v1" and includes several new message types and payload structures.

The extension mechanism operates through capability negotiation, where agents advertise their support for CPMM extensions during the discovery phase. Agents that support CPMM can engage in full economic coordination, while agents that do not support the extension fall back to traditional ACNBP negotiation patterns.

**Extension Advertisement**: During the capability advertisement phase (ACNBP Step 4), agents include an extension manifest that specifies their support for various protocol extensions:

```
{
  "capability_id": "nlp.translation.v2",
  "base_specification": { ... },
  "protocol_extensions": [
   {
     "extension_id": "org.cpmm.economic-coordination.v1",
     "version": "1.0",
     "supported_features": [
       "dynamic_pricing",
       "quality_negotiation",
       "micropayments",
       "sla_enforcement"
     ]
   }
  ]
}
```

**Extension Negotiation**: During the negotiation phases (ACNBP Steps 6-7), agents can include extension-specific payloads that are processed only by agents supporting the corresponding extensions. This allows for graceful degradation when agents have different extension support levels.

### 4.1.2 Economic State Machine Integration

The CPMM economic coordination is integrated into the ACNBP state machine through the addition of economic state transitions and checkpoints. Each ACNBP step is augmented with optional economic processing that occurs in parallel with the standard protocol operations.

**State Augmentation**: The ACNBP state machine is extended with economic state variables:

- **Economic Context**: Current pricing information, payment status, and SLA commitments
- **Payment State**: Escrow status, payment confirmations, and refund capabilities
- **Quality Metrics**: Real-time quality measurements and SLA compliance tracking
- **Reputation State**: Transaction history and reputation scores for participating agents

**Transition Guards**: Economic constraints are added as transition guards that prevent protocol progression when economic conditions are not satisfied. For example, the transition from Negotiate-Response to Bind requires that payment terms have been agreed upon and that necessary payment capabilities are available.

**Rollback Mechanisms**: Economic failures trigger protocol rollback to previous stable states, ensuring that partial economic commitments do not leave the system in inconsistent states.

### 4.1.3 Message Flow and Economic Coordination

The integration preserves the fundamental ACNBP message flow while adding economic coordination at key decision points. The following describes how each ACNBP step is enhanced with economic functionality as shown in Figure 4:

**Step 1-2 (Discover/Pre-screen)**: Enhanced with economic filtering criteria, allowing agents to pre-screen potential partners based on pricing models, payment preferences, and economic reputation.

**Step 3-4 (Negotiate-Request/Response)**: Extended with detailed economic proposals including pricing models, quality-price trade-offs, and payment terms.

Example:
Buyer → Seller: {"quality_request": {"latency": "<50ms", "accuracy": ">98%"},
         "max_price": 0.005, "payment_method": "H402"}
Seller → Buyer: {"price_quote": 0.003, "sla_commitment": "...",
          "economic_proposal": {...}}

**Step 5-6 (Bind/Commit)**: Augmented with cryptographic payment commitments and escrow establishment.

**Step 7-8 (Execute/Verify)**: Enhanced with real-time quality monitoring and automatic payment triggering based on SLA compliance.

**Step 9-10 (Release/Audit)**: Extended with payment settlement, refund processing, and reputation updates.

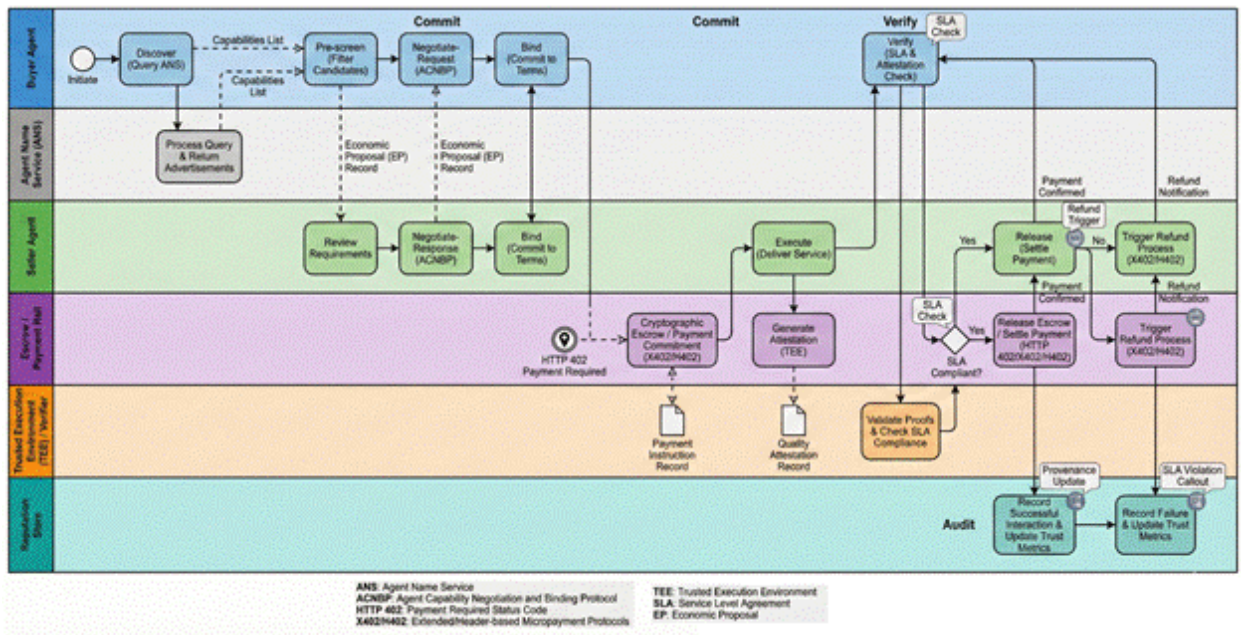

Figure 4 - From service discovery to audit: highlighting economic checkpoints and trust anchors in the Capability-Priced Micro-Markets (CPMM)

## 4.2 Economic Payload Structures

The economic coordination requires rich data structures that capture pricing information, quality specifications, payment terms, and service level agreements. This section specifies the detailed payload structures used in CPMM-enhanced ACNBP messages.

### 4.2.1 Economic Proposal (EP) Record

The Economic Proposal record is the core data structure for communicating economic terms during negotiation. It is included in ACNBP Step 4 (Capability Advertisement) and Step 6 (Negotiate-Response) messages.

```
{
  "economic_proposal": {
    "version": "1.0",
    "proposal_id": "uuid-v4",
    "timestamp": "iso8601-timestamp",
```

```json
"pricing_model": {
  "type": "dynamic_quality_based",
  "base_price": {
    "amount": "decimal",
    "currency": "currency_code",
    "precision": "integer"
  },
  "quality_multipliers": [
    {
      "dimension": "latency",
      "function": "exponential",
      "parameters": {"base": 1.0, "exponent": -0.5}
    },
    {
      "dimension": "accuracy",
      "function": "linear",
      "parameters": {"slope": 2.0, "intercept": 0.0}
    }
  ],
  "volume_discounts": [
    {
      "threshold": 100,
```

```json
      "discount_rate": 0.05
    }
  ]
},
"payment_terms": {
  "accepted_methods": ["X402", "H402", "lightning"],
  "payment_timing": "post_delivery",
  "escrow_required": true,
  "refund_policy": {
    "full_refund_conditions": ["service_failure", "sla_violation"],
    "partial_refund_conditions": ["quality_degradation"],
    "refund_timeframe": "immediate"
  }
},
"service_level_agreement": {
  "quality_guarantees": [
    {
      "dimension": "latency",
      "guarantee": "< 100ms",
      "penalty": "5% price reduction per 10ms over"
    },
    {
```

```json
      "dimension": "accuracy",
      "guarantee": "> 95%",
      "penalty": "full refund if < 90%"
    }
  ],
  "availability_guarantee": "99.9%",
  "capacity_limits": {
    "max_concurrent_requests": 10,
    "max_requests_per_hour": 1000
  }
},
"nanda_capability_hash": "sha256-hash",
"cryptographic_commitment": {
  "commitment_hash": "sha256-hash",
  "signature": "ed25519-signature",
  "public_key": "ed25519-public-key"
 }
 }
}
```

**Pricing Model Specification**: The pricing model supports multiple pricing strategies including fixed pricing, quality-based pricing, volume discounts, and time-based pricing. The modular structure allows for complex pricing schemes while maintaining computational efficiency.

**Payment Terms**: Payment terms specify accepted payment methods, timing requirements, and escrow policies. The support for multiple payment methods (X402, H402, Lightning Network) ensures compatibility with different payment infrastructures.

**Service Level Agreements**: SLAs provide formal guarantees about service quality and availability, with automatic penalty mechanisms for non-compliance. The penalties are designed to align incentives and provide automatic compensation for quality failures.

**Cryptographic Commitment**: All economic proposals include cryptographic commitments that prevent agents from modifying their offers after submission. This ensures the integrity of the negotiation process and prevents strategic manipulation.

### 4.2.2 Payment Instruction Record

Payment instructions are included in ACNBP Step 6 (Commit) messages and specify the detailed payment parameters for the transaction.

```
{
  "payment_instruction": {
    "version": "1.0",
    "instruction_id": "uuid-v4",
    "payment_method": "H402",
    "amount": {
      "base_amount": "decimal",
      "currency": "currency_code",
      "quality_adjustments": [
        {
          "dimension": "latency",
          "measured_value": "85ms",
          "adjustment": "+0.15"
        }
```

```json
  ],
  "final_amount": "decimal"
},
"payment_schedule": {
  "type": "post_delivery",
  "trigger_conditions": ["service_completion", "quality_verification"],
  "timeout": "300s"
},
"escrow_details": {
  "escrow_agent": "nanda_capability_id",
  "escrow_contract": "smart_contract_address",
  "release_conditions": ["mutual_agreement", "sla_compliance", "timeout"]
},
"refund_capability": {
  "capability_token": "nanda_bounded_token",
  "refund_conditions": ["service_failure", "sla_violation"],
  "automatic_triggers": true
},
"cryptographic_proof": {
  "payment_commitment": "sha256-hash",
  "signature": "ed25519-signature",
  "timestamp": "iso8601-timestamp"
```

}

    }

}

**Dynamic Amount Calculation**: The payment amount is calculated dynamically based on actual service quality, allowing for real-time price adjustments that reflect delivered value.

**Escrow Integration**: Payment instructions include detailed escrow specifications that integrate with NANDA's capability-based security model, ensuring that payments are held securely until service delivery is verified.

**Refund Capabilities**: The refund capability mechanism leverages Nanda's bounded tokens to provide automatic refund processing without requiring human intervention.

### 4.2.3 Quality Attestation Record

Quality attestations are generated during ACNBP Step 8 (Verify) and provide cryptographic proof of service quality for payment and reputation purposes.

{

  "quality_attestation": {

    "version": "1.0",

    "attestation_id": "uuid-v4",

    "service_instance_id": "uuid-v4",

    "timestamp": "iso8601-timestamp",

    "quality_measurements": [

      {

        "dimension": "latency",

        "measured_value": "85ms",

        "measurement_method": "client_side_timing",

        "confidence_interval": "±5ms"

```json
      },
      {
        "dimension": "accuracy",
        "measured_value": "97.3%",
        "measurement_method": "reference_comparison",
        "confidence_interval": "±1.2%"
      }
    ],
    "sla_compliance": {
      "overall_compliance": true,
      "violations": [],
      "penalties_applied": []
    },
    "attestation_source": {
      "type": "trusted_execution_environment",
      "tee_identity": "sgx_enclave_id",
      "attestation_report": "base64_encoded_report"
    },
    "cryptographic_proof": {
      "measurement_hash": "sha256-hash",
      "signature": "ed25519-signature",
      "certificate_chain": ["cert1", "cert2", "root_cert"]
```

```
      }
    }
}
```

**Measurement Methodology**: Quality attestations specify the methodology used for quality measurement, enabling other agents to assess the reliability and validity of the measurements.

**TEE Integration**: Attestations generated by trusted execution environments provide strong guarantees about measurement integrity, preventing manipulation by malicious agents.

**Certificate Chains**: The cryptographic proof includes certificate chains that enable verification of attestation authenticity without requiring online connectivity to certificate authorities.

## 4.3 HTTP 402 Integration and Payment Rails

The integration with HTTP 402 and its X402/H402 successors provides the low-level payment infrastructure that enables efficient micropayments within the CPMM framework.

### 4.3.1 HTTP 402 Response Enhancement

Standard HTTP 402 responses are enhanced with CPMM-specific headers that provide rich payment semantics while maintaining compatibility with existing HTTP infrastructure.

**Enhanced 402 Response Headers**:

HTTP/1.1 402 Payment Required

Content-Type: application/json

X402-Payment-Required: amount=0.001 currency=USD method=H402

X402-Payment-Address: H402://payment.endpoint/invoice/12345

X402-Payment-Metadata: sla_vector=latency:100ms,accuracy:95% refund_policy=automatic

CPMM-Economic-Proposal: base64_encoded_ep_record

CPMM-Negotiation-Token: jwt_token_for_continued_negotiation

Cache-Control: no-cache

**Payment Metadata Integration**: The X402-Payment-Metadata header includes CPMM-specific information such as SLA vectors and refund policies, enabling clients to make informed payment decisions.

**Negotiation Continuity**: The CPMM-Negotiation-Token header provides a mechanism for continuing ACNBP negotiation after the initial 402 response, enabling multi-round negotiation within the HTTP request-response paradigm.

### 4.3.2 H402 Micropayment Protocol Integration

The H402 protocol provides cryptographic payment verification directly in HTTP headers, enabling atomic payment and service delivery.

**H402 Payment Headers**:

H402-Payment-Key: ed25519_public_key_base64

H402-Payment-Amount: 0.001

H402-Payment-Currency: USD

H402-Payment-Invoice: invoice_id_12345

H402-Payment-Signature: schnorr_signature_base64

H402-Payment-Timestamp: unix_timestamp

H402-Quality-Request: latency:100ms,accuracy:95%

H402-SLA-Acceptance: sla_hash_sha256

**Atomic Payment Semantics**: The H402 protocol ensures that payments are atomic with respect to service delivery. The payment signature commits to both the payment amount and the quality requirements, ensuring that payments are only valid for services that meet the specified criteria.

**Quality-Conditional Payments**: The H402-Quality-Request header enables payments that are conditional on achieving specified quality levels, providing automatic quality enforcement without requiring separate verification steps.

### 4.3.3 Payment Verification and Settlement

The payment verification process integrates cryptographic verification with quality attestation to ensure that payments are only settled when services meet agreed-upon criteria.

**Verification Algorithm**:

1. **Signature Verification**: Verify the H402 payment signature using the provided public key and payment details.

2. **Quality Verification**: Compare delivered service quality against the quality requirements specified in the payment headers.

3. **SLA Compliance Check**: Verify that the service delivery complies with the accepted SLA terms.

4. **Attestation Validation**: Validate quality attestations using TEE verification and certificate chain validation.

5. **Payment Settlement**: If all verification steps pass, settle the payment according to the specified payment method.

**Settlement Integration**: The settlement process integrates with various payment backends including cryptocurrency networks, traditional payment processors, and internal accounting systems. The modular design allows for easy integration of new payment methods as they become available.

## 4.4 Refund and Dispute Resolution Mechanisms

Automated refund and dispute resolution are critical for maintaining trust in the CPMM ecosystem, particularly given the autonomous nature of agent interactions and the potential for service failures.

### 4.4.1 Automatic Refund Triggers

The CPMM framework implements several automatic refund triggers that provide immediate compensation for service failures without requiring human intervention.

**Service Failure Refunds**: Complete service failures (no response, system errors, capability unavailability) trigger immediate full refunds. The refund process is initiated automatically by the escrow system when service delivery timeouts are exceeded or when error conditions are detected.

**SLA Violation Refunds**: Violations of service level agreements trigger automatic partial or full refunds based on the severity of the violation and the penalty structure specified in the economic proposal. The refund calculation is performed automatically based on quality attestations and SLA terms.

**Quality Degradation Refunds**: Significant quality degradation below agreed-upon thresholds triggers automatic partial refunds. The refund amount is calculated based on the quality-price relationship specified in the economic proposal.

### 4.4.2 NANDA-Based Refund Capabilities

The refund mechanism leverages NANDA's capability-based security model to provide secure, automated refund processing.

**Refund Capability Tokens**: When a payment is made, the buyer receives a refund capability token that represents the right to claim a refund under specified conditions. This token is implemented as a Nanda bounded capability that can only be exercised under the conditions specified in the economic proposal.

**Smart Contract Integration**: Refund capabilities are backed by smart contracts that automatically verify refund conditions and release funds when appropriate. The smart contracts integrate with quality attestation systems to automatically detect SLA violations and service failures.

**Cryptographic Verification**: All refund claims are cryptographically verified to ensure that only legitimate refunds are processed. The verification process includes checking the refund capability token, validating quality attestations, and confirming SLA violation conditions.

### 4.4.3 Escalation and Human Arbitration

While the CPMM framework is designed to handle most disputes automatically, some complex cases may require human arbitration. The system provides escalation mechanisms for cases that cannot be resolved automatically.

**Escalation Triggers**: Disputes are escalated to human arbitration when automatic resolution mechanisms fail, when agents explicitly request human review, or when the dispute involves complex interpretation of contract terms that cannot be automatically verified.

**Arbitration Protocol**: The arbitration protocol leverages the existing ACNBP audit trail to provide arbitrators with complete transaction history and evidence. Arbitrators have access to all negotiation messages, quality attestations, and payment records.

**Binding Arbitration**: Arbitration decisions are implemented through the same cryptographic mechanisms used for automatic dispute resolution, ensuring that arbitration outcomes are enforceable and cannot be ignored by participating agents.

## 4.5 Implementation Architecture and Deployment Considerations

This section describes the practical implementation architecture for CPMM and addresses key deployment considerations for real-world agent systems.

### 4.5.1 Layered Architecture Design

The CPMM implementation follows a layered architecture that separates economic logic from protocol mechanics, enabling modular development and deployment.

**Protocol Layer**: The lowest layer implements the core ACNBP protocol with CPMM extensions. This layer handles message serialization, cryptographic operations, and network communication.

**Economic Engine Layer**: The middle layer implements the economic logic including pricing algorithms, quality evaluation, and payment processing. This layer is designed to be pluggable, allowing different economic strategies to be implemented and tested.

**Application Layer**: The highest layer provides APIs and interfaces for agent applications to interact with the CPMM framework. This layer abstracts the complexity of economic coordination and provides simple interfaces for common operations.

**Integration Points**: The architecture includes well-defined integration points for external systems including payment processors, quality measurement tools, and reputation systems.

### 4.5.2 Scalability and Performance Considerations

The CPMM framework is designed to scale to large numbers of agents and high transaction volumes while maintaining low latency and high reliability.

**Distributed Processing**: Economic calculations are distributed across multiple nodes to avoid bottlenecks and single points of failure. The distributed design ensures that the system can handle high transaction volumes without degrading performance.

**Caching and Optimization**: Frequently accessed data such as capability advertisements and reputation scores are cached to reduce latency. The caching system is designed to maintain consistency while providing fast access to critical information.

**Asynchronous Processing**: Long-running operations such as quality verification and payment settlement are processed asynchronously to avoid blocking agent interactions. The asynchronous design ensures that agents can continue operating even when some background processes are delayed.

### 4.5.3 Security and Privacy Implementation

The implementation includes comprehensive security and privacy protections that go beyond the basic cryptographic primitives to address practical deployment concerns.

**Key Management**: The system includes secure key management for agent identities, payment credentials, and capability certificates. Key rotation and revocation mechanisms ensure that compromised keys can be quickly replaced without disrupting system operation.

**Privacy Protection**: Agent interactions are designed to minimize information disclosure while still enabling effective economic coordination. Techniques such as zero-knowledge proofs and secure multi-party computation are used where appropriate to protect sensitive information.

**Audit and Compliance**: The system maintains comprehensive audit trails that enable compliance with regulatory requirements while protecting agent privacy. The audit system is designed to provide accountability without compromising the autonomy of agent operations.

This detailed protocol integration specification provides the foundation for implementing CPMM in real-world agent systems. The integration with ACNBP ensures compatibility with existing agent infrastructure while the economic extensions enable sophisticated market mechanisms. The payment integration provides efficient micropayment capabilities, and the dispute resolution mechanisms ensure that the system can operate reliably even in the presence of failures and conflicts.

## 4.6 Failure Mode Analysis and Robustness

### 4.6.1 TEE Compromise Scenarios
- When hardware attestations become unreliable
- Fallback to multi-party verification mechanisms
- Economic incentives under degraded trust

### 4.6.2 Network Partition Tolerance
- Protocol behavior during network splits
- Autonomous operation with limited connectivity
- Reconciliation strategies for rejoining agents

### 4.6.3 Economic Attack Recovery
- Detection mechanisms for coordinated manipulation
- Automatic market circuit breakers
- Recovery protocols for compromised price discovery

# 5. Privacy Economics and Security Analysis

The CPMM framework operates in an environment where agents must balance the benefits of information sharing with the costs of privacy loss. This section develops a formal theory of privacy economics in agent markets and provides comprehensive security analysis of the system's resistance to various attack vectors.

## 5.1 Privacy Elasticity Theory

Privacy in agent systems is not a binary property but rather a continuous variable that agents can control through their disclosure decisions. We develop a formal framework for analyzing how privacy choices affect market outcomes and agent welfare.

### 5.1.1 Information Disclosure Model

We model an agent's information disclosure decision as a choice of disclosure level $\sigma \in [0, 1]$, where $\sigma = 0$ represents complete privacy (no information disclosed) and $\sigma = 1$ represents complete transparency (all information disclosed). The disclosure level affects both the agent's market opportunities and its privacy costs.

Let $G(\sigma)$ represent the agent's capability graph, where higher disclosure levels reveal more detailed information about the agent's capabilities, performance history, and internal state. The disclosed graph $G(\sigma)$ is a subgraph of the agent's complete capability graph $G(1)$, with the property that $G(\sigma_1) \subseteq G(\sigma_2)$ for $\sigma_1 \leq \sigma_2$.

The agent's disclosure decision affects its market outcomes through several channels:

**Market Access**: Higher disclosure levels enable access to more trading opportunities, as other agents can better assess compatibility and trustworthiness.

**Price Premium**: Agents with verifiable capability information may command higher prices due to reduced uncertainty about service quality.

**Competitive Advantage**: Strategic information disclosure can provide competitive advantages by revealing strengths while concealing weaknesses.

**Privacy Costs**: Information disclosure imposes privacy costs including increased vulnerability to competitive intelligence gathering and potential security risks.

### 5.1.2 Privacy Cost Function

We model privacy costs using a convex function $C(\sigma)$ that captures the increasing marginal cost of information disclosure. The convexity reflects the fact that initial disclosures (basic capability information) have low privacy costs, while detailed disclosures (internal algorithms, performance metrics, customer lists) have high privacy costs.

A canonical specification is:

$$C(\sigma) = \gamma \cdot \sigma^\alpha$$

where $\gamma > 0$ represents the agent's privacy sensitivity and $\alpha > 1$ captures the convexity of privacy costs. Higher values of $\alpha$ indicate that privacy costs increase rapidly with disclosure level.

The privacy cost function may also depend on market conditions and competitive environment:

$$C(\sigma, \theta) = \gamma(\theta) \cdot \sigma^{\alpha(\theta)}$$

where $\theta$ represents market state variables such as competitive intensity, regulatory environment, and security threat levels.

### 5.1.3 Market Value of Information

The market value of disclosed information depends on how it affects other agents' willingness to trade and the prices they are willing to pay. We model this through a market value function $V(\sigma)$ that captures the expected increase in trading revenue from disclosure level $\sigma$.

The market value function exhibits several important properties:

**Increasing Returns**: Initial disclosures have high marginal value as they enable basic market participation: $V'(\sigma)$ is high for small $\sigma$.

**Diminishing Returns**: Additional disclosures beyond a threshold provide decreasing marginal value: $V''(\sigma) < 0$ for large $\sigma$.

**Network Effects**: The value of disclosure may depend on the disclosure levels of other agents in the market.

A flexible specification that captures these properties is:

$$V(\sigma) = \beta \cdot \sigma / (1 + \delta \cdot \sigma)$$

where $\beta$ represents the maximum potential value of full disclosure and $\delta$ controls the rate of diminishing returns.

### 5.1.4 Optimal Disclosure Decision

The agent's optimal disclosure level $\sigma^*$ solves the optimization problem:

$$\text{maximize } V(\sigma) - C(\sigma)$$

The first-order condition is:

$$V'(\sigma^*) = C'(\sigma^*)$$

This condition states that the optimal disclosure level equates the marginal value of information disclosure with its marginal privacy cost.

**Theorem 5.1 (Existence and Uniqueness)**: Under the assumptions that $V(\sigma)$ is concave and $C(\sigma)$ is convex, there exists a unique optimal disclosure level $\sigma^* \in (0, 1)$.

**Proof**: The objective function $V(\sigma) - C(\sigma)$ is strictly concave due to the concavity of $V$ and convexity of $C$. Since the domain $[0, 1]$ is compact, the maximum exists. Strict concavity ensures uniqueness.

### 5.1.5 Privacy Elasticity of Demand

We define the privacy elasticity of demand as the percentage change in market price resulting from a percentage change in disclosure level:

$$\xi = (\partial p/\partial \sigma) \cdot (\sigma/p)$$

This elasticity measure captures how sensitive market prices are to changes in information disclosure.

**Theorem 5.2 (Privacy Elasticity Bounds)**: Under reasonable assumptions about market structure and agent preferences, the privacy elasticity satisfies:

$$-1 \leq \xi \leq 0$$

The negative elasticity indicates that increased disclosure generally leads to lower prices, as reduced uncertainty decreases the risk premium that buyers must pay.

We further argued the following:

**Selective Disclosure**: Agents benefit from mechanisms that allow selective disclosure of information to different trading partners.

**Privacy Insurance**: Market mechanisms for privacy insurance can help agents manage the trade-off between disclosure and privacy costs.

## 5.2 Information-Theoretic Security Analysis

The security of the CPMM framework depends on the information-theoretic properties of the cryptographic protocols and the economic incentives they create. This section analyzes the system's resistance to various information-theoretic attacks.

### 5.2.1 Capability Verification Security

The capability-based security model relies on cryptographic proofs that agents possess claimed capabilities. We analyze the security of this verification process against various attack strategies.

**Capability Forgery Resistance**: The security against capability forgery depends on the strength of the underlying cryptographic primitives and the structure of the capability delegation chains.

Let $\kappa$ be the security parameter (e.g., key length), and let $\text{AdvForge}(A, \kappa)$ be the advantage of adversary A in forging a valid capability certificate. Under standard cryptographic assumptions:

$$\text{AdvForge}(A, \kappa) \leq \text{negl}(\kappa)$$

where negl(κ) is a negligible function in κ.

**Delegation Chain Verification**: Capability delegation chains must be verifiable without revealing unnecessary information about intermediate delegators. We use zero-knowledge proofs to enable verification while preserving privacy.

The verification process has computational complexity $O(d \cdot \log n)$ where d is the delegation chain depth and n is the number of potential delegators. This logarithmic dependence ensures scalability even in large agent networks.

### 5.2.2 Payment Security Analysis

The micropayment protocols must resist various attacks including double-spending, payment forgery, and transaction malleability.

**Double-Spending Resistance**: The H402 protocol prevents double-spending through the use of unique payment identifiers and cryptographic commitments. Each payment includes a unique nonce that prevents replay attacks.

**Theorem 5.3 (Double-Spending Security)**: Under the discrete logarithm assumption, the probability that an adversary can successfully double-spend a payment is bounded by:

$$P(\text{double-spend}) \leq 2^{-\kappa} + \text{AdvDL}(A, \kappa)$$

where $\text{AdvDL}(A, \kappa)$ is the adversary's advantage in solving the discrete logarithm problem.

**Payment Atomicity**: The payment protocol ensures atomicity between payment and service delivery through cryptographic escrow mechanisms. Payments are only released when service delivery is cryptographically verified.

**Transaction Privacy**: Payment transactions reveal minimal information about agent identities and transaction patterns. The use of ephemeral keys and unlinkable payment identifiers provides strong privacy protection.

### 5.2.3 Quality Attestation Security

Quality attestations must be unforgeable and verifiable while protecting sensitive information about agent capabilities and performance.

**Attestation Integrity**: Quality attestations are generated by trusted execution environments (TEEs) that provide hardware-based security guarantees. The integrity of attestations depends on the security of the underlying TEE technology.

**Theorem 5.4 (Attestation Unforgeability)**: Assuming the security of the TEE platform, the probability that an adversary can forge a valid quality attestation is bounded by:

P(forge-attestation) ≤ AdvTEE(A) + $2^{-\kappa}$

where AdvTEE(A) is the adversary's advantage in compromising the TEE platform.

**Privacy-Preserving Attestations**: Attestations can be generated in a privacy-preserving manner using techniques such as selective disclosure and zero-knowledge proofs. This allows agents to prove quality properties without revealing detailed performance metrics.

## 5.3 Economic Security and Mechanism Design

Beyond cryptographic security, the CPMM framework must be secure against economic attacks that exploit the incentive structure of the market mechanism.

### 5.3.1 Sybil Attack Resistance

Sybil attacks involve creating multiple fake identities to manipulate market outcomes. The capability-based security model provides natural resistance to Sybil attacks by requiring each identity to possess verifiable capabilities.

**Capability Scarcity**: The resistance to Sybil attacks depends on the scarcity of capabilities. If capabilities are easy to obtain or replicate, Sybil attacks become more feasible.

Let $C(k)$ be the cost of obtaining $k$ valid capability certificates, and let $B(k)$ be the benefit from controlling $k$ identities in the market. Sybil attacks are economically infeasible when $C(k) > B(k)$ for all $k > 1$.

**Theorem 5.5 (Sybil Resistance Bound)**: Under the assumption that capabilities have positive acquisition costs, the benefit from Sybil attacks is bounded by:

$$B(k) \leq O(\log k / \sqrt{k})$$

This sublinear bound ensures that the marginal benefit of additional Sybil identities decreases rapidly, making large-scale attacks economically unattractive.

**Proof Sketch**: The benefit from Sybil attacks comes from the ability to manipulate market prices or gain unfair advantages in matching. However, the decentralized nature of the matching process and the requirement for verifiable capabilities limits the effectiveness of such manipulation. The logarithmic factor comes from information-theoretic limits on market manipulation, while the square root factor reflects the diminishing returns to scale in market influence.

### 5.3.2 Market Manipulation Resistance

Market manipulation involves coordinated actions to artificially influence prices or market outcomes. The CPMM framework includes several mechanisms to detect and prevent market manipulation.

**Price Manipulation Detection**: The system monitors for unusual price patterns that may indicate manipulation attempts. Statistical tests are used to detect deviations from expected price behavior.

**Collusion Resistance**: The mechanism design includes features that make collusion difficult and unprofitable. The use of sealed-bid auctions and randomized matching reduces the ability of colluding agents to coordinate their actions.

**Reputation-Based Penalties**: Agents detected engaging in market manipulation face reputation penalties that reduce their future market opportunities. The reputation system is designed to be robust against manipulation attempts.

### 5.3.3 Quality Manipulation Resistance

Agents may attempt to manipulate quality measurements to receive payments for substandard services. The CPMM framework includes several mechanisms to prevent and detect quality manipulation.

**Multi-Source Verification**: Quality measurements are verified using multiple independent sources, making it difficult for agents to manipulate all verification sources simultaneously.

**Cryptographic Commitments**: Agents must commit to quality levels before service delivery, preventing them from adjusting quality claims based on actual performance.

**Incentive Alignment**: The payment structure is designed to align incentives so that providing high-quality service is more profitable than attempting to manipulate quality measurements.

## 5.4 Privacy-Preserving Market Mechanisms

This section describes specific mechanisms that enable market participation while preserving agent privacy.

### 5.4.1 Zero-Knowledge Capability Proofs

Agents can prove possession of capabilities without revealing detailed information about their implementation or performance characteristics.

**Capability Proof Protocol**:

1. The agent generates a zero-knowledge proof $\pi$ that it possesses capability $c$
2. The proof $\pi$ can be verified by other agents without learning details about $c$

3. The proof includes commitments to quality guarantees that can be verified during service delivery

**Theorem 5.6 (Zero-Knowledge Property)**: The capability proof protocol is zero-knowledge, meaning that verifiers learn nothing about the capability beyond its existence and committed properties.

**Efficiency**: The proof generation and verification have computational complexity $O(|c|)$ where $|c|$ is the size of the capability description. This linear complexity ensures practical efficiency even for complex capabilities.

### 5.4.2 Private Information Retrieval for Discovery

The agent discovery process can be implemented using private information retrieval (PIR) techniques that allow agents to search for capabilities without revealing their search queries.

**PIR-Based Discovery Protocol**:

1. The searching agent constructs a PIR query that encodes its capability requirements
2. The ANS processes the query and returns matching capabilities without learning the search criteria
3. The response includes only the information necessary for the searching agent to evaluate potential matches

**Privacy Guarantees**: The PIR protocol ensures that the ANS learns nothing about the searching agent's requirements, while the searching agent learns only about capabilities that match its criteria.

**Communication Complexity**: The PIR protocol has communication complexity $O(\sqrt{n})$ where $n$ is the number of registered capabilities, providing significant efficiency improvements over naive approaches.

### 5.4.3 Secure Multi-Party Computation for Auctions

Complex auction mechanisms can be implemented using secure multi-party computation (SMC) techniques that enable price discovery without revealing individual bids.

**SMC Auction Protocol**:

1. Agents submit encrypted bids to a set of auction servers
2. The servers jointly compute the auction outcome without learning individual bid values
3. The auction result is revealed while keeping individual bids private

**Security Properties**: The SMC auction protocol provides:

- **Bid Privacy**: Individual bids remain private even if a subset of auction servers are compromised
- **Outcome Correctness**: The auction outcome is computed correctly even in the presence of malicious participants
- **Fairness**: All participants receive the auction outcome simultaneously, preventing early information leakage

## 5.5 Formal Security Model and Analysis

This section presents a formal security model for the CPMM framework and analyzes its security properties under various threat scenarios.

### 5.5.1 Adversarial Model

We consider a powerful adversary with the following capabilities:

**Network Control**: The adversary can intercept, delay, modify, or drop messages between agents. However, the adversary cannot break cryptographic primitives or forge digital signatures.

**Agent Compromise**: The adversary can compromise a subset of agents, gaining access to their private keys and internal state. We assume that the fraction of compromised agents is bounded by some threshold $\tau < 1/2$.

**Economic Resources**: The adversary has significant economic resources and can participate in markets to manipulate prices or outcomes.

**Computational Power**: The adversary has polynomial computational power but cannot solve cryptographically hard problems such as discrete logarithm or factoring.

### 5.5.2 Security Properties

We define several security properties that the CPMM framework should satisfy:

**Capability Authenticity**: Only agents that actually possess a capability can prove possession of that capability.

**Payment Integrity**: Payments are only settled when services are delivered according to agreed-upon terms.

**Quality Verifiability**: Quality claims can be verified by independent parties without revealing sensitive information.

**Privacy Preservation**: Agent interactions reveal minimal information beyond what is necessary for market operation.

**Market Integrity**: Market outcomes reflect genuine supply and demand rather than manipulation attempts.

### 5.5.3 Security Theorems

**Theorem 5.7 (Capability Authenticity)**: Under standard cryptographic assumptions, the probability that an adversary can prove possession of a capability it does not actually possess is negligible in the security parameter.

**Theorem 5.8 (Payment Integrity)**: Under the assumption that fewer than half of the escrow agents are compromised, payments are settled correctly with probability $1 - \text{negl}(\kappa)$.

**Theorem 5.9 (Quality Verifiability)**: Quality attestations generated by secure TEEs are unforgeable with probability $1 - \text{negl}(\kappa)$, assuming the security of the underlying TEE platform.

**Theorem 5.10 (Privacy Preservation)**: The CPMM protocols satisfy differential privacy with parameter $\varepsilon$, meaning that the participation of any individual agent has minimal impact on the information revealed about other agents.

**Theorem 5.11 (Market Integrity)**: Under the assumption that fewer than $\tau$ fraction of agents are adversarial, market outcomes converge to competitive equilibrium with probability $1 - \delta$ for any $\delta > 0$.

### 5.5.4 Security Analysis Under Composition

The CPMM framework consists of multiple interacting protocols, and its overall security depends on how these protocols compose. We analyze the security of the composed system using the Universal Composability (UC) framework.

**UC Security**: We prove that the CPMM protocols are UC-secure, meaning that they remain secure when composed with arbitrary other protocols. This provides strong guarantees about the security of the overall system.

**Theorem 5.12 (UC Security)**: The CPMM protocols UC-realize their ideal functionalities in the presence of adaptive adversaries that can compromise agents during protocol execution.

This UC security guarantee ensures that the CPMM framework can be safely deployed in complex environments where multiple protocols are running simultaneously.

## 5.6 Privacy-Utility Trade-offs and Optimization

The final component of our privacy analysis examines the fundamental trade-offs between privacy and utility in agent markets and provides optimization frameworks for managing these trade-offs.

### 5.6.1 Privacy-Utility Frontier

We model the privacy-utility trade-off using a frontier that characterizes the maximum utility achievable for any given level of privacy protection.

Let $U(\sigma, \theta)$ be the utility function that depends on disclosure level $\sigma$ and market conditions $\theta$. Let $P(\sigma)$ be the privacy level, which decreases with disclosure. The privacy-utility frontier is defined as:

$$F(p) = \max\{U(\sigma, \theta) : P(\sigma) \geq p\}$$

The frontier $F(p)$ is typically concave, reflecting the fact that achieving high privacy levels requires sacrificing significant utility.

**Theorem 5.13 (Frontier Properties)**: Under reasonable assumptions about utility and privacy functions, the privacy-utility frontier is:

1. Decreasing: $F'(p) \leq 0$ (higher privacy requires lower utility)
2. Concave: $F''(p) \leq 0$ (diminishing returns to privacy protection)
3. Continuous: $F(p)$ is continuous in p

### 5.6.2 Dynamic Privacy Optimization

In dynamic environments, agents must continuously adjust their privacy-utility trade-offs based on changing market conditions and threat levels.

We formulate this as a dynamic optimization problem:

$$\text{maximize } \int_0^T e^{-rt} [U(\sigma_t, \theta_t) - C(\sigma_t)] \, dt$$

subject to privacy constraints and market dynamics.

The solution involves a feedback control policy $\sigma_t = \pi(\theta_t, P_t)$ where $P_t$ represents the agent's current privacy state.

**Theorem 5.14 (Optimal Privacy Policy)**: The optimal privacy policy has a threshold structure: there exists a threshold function $\theta^*(P)$ such that agents increase disclosure when $\theta_t > \theta^*(P_t)$ and decrease disclosure when $\theta_t < \theta^*(P_t)$.

This threshold structure provides practical guidance for implementing adaptive privacy policies in agent systems.

The comprehensive privacy and security analysis presented in this section demonstrates that the CPMM framework can provide strong security guarantees while enabling flexible privacy management. The privacy elasticity theory provides insights into the economic value of privacy, while the security analysis shows that the system is robust against a wide range of attacks. The privacy-preserving mechanisms enable agents to participate in markets while maintaining control over their information disclosure, and the formal security model provides rigorous guarantees about system behavior under adversarial conditions.

## 6. Related Work

The CPMM framework builds upon and extends several decades of research in multi-agent systems, mechanism design, distributed computing economics, and cryptographic protocols. This section provides a comprehensive survey of related work and positions CPMM within the broader landscape of agent coordination and economic mechanisms.

### 6.1 Agent Communication and Coordination Protocols

The foundation of multi-agent coordination lies in communication protocols that enable agents to discover, negotiate, and coordinate their activities. The evolution of these protocols has progressed from simple task allocation mechanisms to sophisticated frameworks supporting complex economic interactions.

#### 6.1.1 Contract Net Protocol and Extensions

The Contract Net Protocol (CNP), introduced by Smith in 1980 [14], established the foundational paradigm for distributed task allocation in multi-agent systems. CNP operates through a bidding mechanism where manager agents announce tasks and contractor agents submit bids for execution. The protocol's simplicity and effectiveness led to widespread adoption and numerous extensions.

The original CNP suffers from several limitations that CPMM addresses. First, CNP assumes static task descriptions and does not support dynamic quality negotiation or complex pricing models. Second, the protocol lacks mechanisms for capability verification, making it vulnerable to agents that bid on tasks they cannot actually perform. Third, CNP does not include payment or settlement mechanisms, assuming that task completion is sufficient compensation.

Several extensions to CNP have addressed specific limitations. The Norm-based CNP [15] introduces social norms to improve coordination efficiency, while the Multi-round CNP [16] enables iterative negotiation for complex tasks. However, these extensions still lack the comprehensive economic framework and cryptographic security guarantees provided by CPMM.

Recent work on market-based approaches to multi-agent coordination [17] has explored the use of auction mechanisms for task allocation. These approaches recognize the importance of economic incentives but typically assume centralized auctioneers and do not address the challenges of decentralized capability verification and micropayment settlement that are central to CPMM.

### 6.1.2 FIPA Standards and Agent Communication Languages

The Foundation for Intelligent Physical Agents (FIPA) developed comprehensive standards for agent communication, including the Agent Communication Language (ACL) and interaction protocols [18]. FIPA ACL provides a rich semantic framework for agent communication based on speech act theory, enabling agents to express complex intentions and commitments.

FIPA's interaction protocols, including the Contract Net Interaction Protocol and various auction protocols, provide standardized patterns for common coordination tasks. These protocols have been widely adopted in academic and commercial multi-agent systems, providing a foundation for interoperability.

However, FIPA standards focus primarily on communication semantics and do not address economic coordination or payment mechanisms. The protocols assume that agents will honor their commitments without providing enforcement mechanisms or economic incentives. CPMM extends the FIPA approach by integrating economic mechanisms directly into the communication protocols while maintaining compatibility with existing FIPA-based systems.

### 6.1.3 Modern Agent Frameworks

Recent developments in agent frameworks have focused on practical deployment and integration with existing systems. The Model Context Protocol (MCP) [1] provides standardized interfaces for AI agents to interact with tools and resources, while Agent-to-Agent (A2A) frameworks [2] enable direct agent communication without centralized coordination.

Cisco's Agent Gateway Protocol (AGP) [3] addresses enterprise deployment concerns by providing secure, scalable agent communication within corporate networks. These frameworks represent significant advances in practical agent deployment but generally assume static pricing models or delegate economic coordination to external systems.

CPMM complements these frameworks by providing the missing economic layer that enables autonomous agent commerce. The integration with ACNBP ensures compatibility with existing agent communication patterns while adding sophisticated economic capabilities.

## 6.2 Mechanism Design and Auction Theory

The theoretical foundation for CPMM lies in mechanism design theory, which studies how to design rules and institutions that produce desired outcomes when participants have private information and strategic incentives.

### 6.2.1 Classical Mechanism Design

The seminal work of Vickrey [19], Clarke [20], and Groves [21] established the theoretical foundations of mechanism design through the development of the VCG mechanism. VCG auctions achieve truthful bidding and efficient allocation by charging bidders their externality on other participants. This work demonstrated that well-designed mechanisms can align individual incentives with social objectives.

The revelation principle, established by Myerson [22], shows that any mechanism can be implemented through a direct mechanism where agents truthfully report their private information. This principle provides a powerful tool for mechanism design but assumes the existence of a trusted mechanism designer who can enforce payments and commitments.

CPMM extends classical mechanism design to decentralized environments where no single party can enforce mechanism rules. The use of cryptographic protocols and capability-based security provides the enforcement mechanisms that enable truthful behavior without relying on trusted third parties.

### 6.2.2 Combinatorial Auctions and Complex Valuations

The extension of auction theory to handle complex valuations and combinatorial bidding has been a major focus of research. Combinatorial auctions allow bidders to submit bids on packages of items, enabling more efficient allocation when items are complements or substitutes [23].

The computational complexity of combinatorial auctions has been extensively studied, with the winner determination problem being NP-hard in general [24]. Various approximation algorithms and heuristics have been developed to make combinatorial auctions practical for real-world applications.

CPMM incorporates combinatorial auction mechanisms for workflow composition, where agents can bid on subsets of tasks within complex workflows. The integration with capability-based security ensures that only agents with verified capabilities can participate in auctions, reducing the complexity of winner determination by eliminating infeasible bids.

### 6.2.3 Dynamic and Repeated Mechanisms

The analysis of dynamic mechanisms, where the same agents interact repeatedly over time, has revealed important insights about learning, reputation, and long-term incentives. Repeated interaction enables the use of reputation mechanisms and the threat of future punishment to enforce cooperation [25].

Dynamic mechanism design faces the challenge of balancing current efficiency with future incentives. Agents may have incentives to manipulate their current behavior to influence future mechanism outcomes, leading to complex strategic considerations.

CPMM addresses these challenges through the integration of reputation systems with economic mechanisms. The capability-based security model provides a foundation for persistent agent identity, enabling reputation accumulation while preserving privacy through selective disclosure mechanisms.

## 6.3 Distributed Systems Economics

The application of economic principles to distributed systems has emerged as a powerful approach for resource allocation and coordination in large-scale systems.

### 6.3.1 Computational Economics

The field of computational economics applies economic models to computational problems, particularly in distributed systems where resources must be allocated among competing users [26]. Early work focused on CPU scheduling and network bandwidth allocation using market mechanisms.

The Spawn system [27] demonstrated the use of economic mechanisms for distributed computation, allowing users to bid for computational resources. Similar approaches have been applied to grid computing [28] and cloud computing [29], where economic mechanisms provide efficient resource allocation in heterogeneous environments.

CPMM extends computational economics to the domain of AI agent services, where the "resources" being allocated are complex capabilities rather than simple computational resources. The quality-dependent pricing models and SLA enforcement mechanisms address the unique challenges of service-oriented resource allocation.

### 6.3.2 Peer-to-Peer Economics

Peer-to-peer systems have explored various economic mechanisms for incentivizing participation and preventing free-riding. BitTorrent's tit-for-tat mechanism [30] provides a simple but effective approach to encouraging sharing, while more sophisticated reputation systems have been developed for other P2P applications [31].

The challenge in P2P economics is providing incentives for participation while maintaining the decentralized nature of the system. Centralized payment systems conflict with the P2P philosophy, leading to the development of various token-based and reputation-based mechanisms.

CPMM addresses P2P economics challenges through the integration of cryptographic micropayments with decentralized capability verification. The system maintains the decentralized nature of P2P systems while enabling efficient economic coordination.

### 6.3.3 Blockchain and Cryptocurrency Economics

The emergence of blockchain technology and cryptocurrencies has enabled new forms of decentralized economic coordination. Smart contracts provide programmable enforcement of economic agreements, while cryptocurrencies enable real time payments without traditional financial intermediaries [32].

Various blockchain-based systems have explored agent coordination and economic mechanisms. The Ethereum platform [33] enables complex smart contracts that can implement sophisticated auction mechanisms, while specialized blockchain platforms have been developed for specific application domains.

CPMM leverages blockchain technology for payment settlement and smart contract enforcement while avoiding the scalability and energy consumption issues of traditional blockchain systems. The integration with HTTP 402 and X402/H402 protocols provides efficient micropayment capabilities that are compatible with existing web infrastructure.

## 6.4 Privacy-Preserving Mechanisms

The integration of privacy protection with economic mechanisms has become increasingly important as concerns about data privacy and surveillance have grown.

### 6.4.1 Differential Privacy in Mechanism Design

Differential privacy [34] provides a rigorous framework for quantifying and limiting privacy loss in data analysis. The application of differential privacy to mechanism design has led to the development of privacy-preserving auctions and other economic mechanisms [35].

The challenge in differentially private mechanism design is balancing privacy protection with mechanism efficiency. Strong privacy guarantees typically require adding noise to mechanism outcomes, which can reduce efficiency and revenue.

CPMM incorporates differential privacy guarantees through the selective disclosure mechanisms and zero-knowledge proofs. The privacy elasticity analysis provides a framework for understanding the trade-offs between privacy and economic efficiency.

### 6.4.2 Secure Multi-Party Computation

Secure multi-party computation (SMC) enables multiple parties to jointly compute functions over their private inputs without revealing the inputs to each other [36]. SMC has been applied to various economic mechanisms, including auctions and voting systems.

The computational overhead of SMC has historically limited its practical application, but recent advances in cryptographic protocols have made SMC increasingly practical for real-world applications [37].

CPMM uses SMC techniques for privacy-preserving auction mechanisms and capability verification. The modular design allows for the integration of different SMC protocols based on the specific privacy and efficiency requirements of different applications.

### 6.4.3 Zero-Knowledge Proofs

Zero-knowledge proofs enable one party to prove knowledge of information without revealing the information itself [38]. These proofs have found applications in various privacy-preserving systems, including cryptocurrency protocols and identity verification systems.

Recent advances in zero-knowledge proof systems, particularly zk-SNARKs [39] and zk-STARKs [40], have made these techniques practical for complex applications. The development of general-purpose zero-knowledge virtual machines has further expanded the applicability of these techniques.

CPMM integrates zero-knowledge proofs for capability verification and quality attestation. The proofs enable agents to demonstrate their capabilities and performance without revealing sensitive implementation details or competitive information.

## 6.5 Micropayment Systems and Digital Cash

The vision of frictionless micropayments has been a long-standing goal in digital commerce, with various technical and economic challenges hindering widespread adoption.

### 6.5.1 Early Digital Cash Systems

Early digital cash systems, including DigiCash [41] and Millicent [42], explored various approaches to enabling small-value digital transactions. These systems faced challenges including double-spending prevention, scalability, and merchant adoption.

The failure of early digital cash systems was attributed to various factors including technical limitations, regulatory concerns, and the lack of compelling use cases. The emergence of the web and e-commerce created demand for micropayments, but existing payment systems were not well-suited to small-value transactions.

### 6.5.2 Cryptocurrency and Lightning Networks

The development of Bitcoin [43] and other cryptocurrencies provided new approaches to digital payments, including the potential for micropayments. However, the scalability limitations and transaction fees of traditional blockchain systems made micropayments impractical.

Layer-2 solutions, particularly the Lightning Network [44], address these limitations by enabling off-chain micropayments that are settled periodically on the blockchain. These systems provide the efficiency needed for micropayments while maintaining the security guarantees of the underlying blockchain.

CPMM integrates with various payment systems including Lightning Network and other cryptocurrency-based micropayment solutions. The modular payment architecture allows for the integration of new payment methods as they become available.

### 6.5.3 HTTP 402 and Web Micropayments

The HTTP 402 "Payment Required" status code was included in the original HTTP specification but remained largely unused due to the lack of practical micropayment systems. Recent developments in digital payments have renewed interest in HTTP 402 as a foundation for web-based micropayments [45].

Various extensions to HTTP 402 have been proposed, including the X402 and H402 protocols that add cryptographic payment verification directly to HTTP headers. These protocols enable efficient micropayments without requiring modifications to existing web infrastructure.

CPMM builds upon these HTTP 402 extensions to provide seamless integration of economic mechanisms with web-based agent communication. The integration ensures compatibility with existing web infrastructure while enabling sophisticated economic coordination.

## 6.6 Trust and Reputation Systems

Trust and reputation systems provide mechanisms for establishing and maintaining trust in environments where participants may be unknown or have conflicting interests.

### 6.6.1 Centralized Reputation Systems

Centralized reputation systems, such as those used by eBay [46] and Amazon [47], aggregate feedback from multiple transactions to provide reputation scores for participants. These systems have been successful in enabling trust in online marketplaces but suffer from various limitations including manipulation vulnerability and centralized control.

The design of effective reputation systems involves balancing various trade-offs including accuracy, manipulation resistance, and privacy protection. Various aggregation mechanisms and incentive structures have been proposed to address these challenges [48].

### 6.6.2 Decentralized Trust Systems

Decentralized trust systems aim to provide reputation and trust mechanisms without relying on centralized authorities. These systems face additional challenges including Sybil attacks, collusion, and the lack of global coordination.

Web-of-trust systems, such as those used in PGP [49], provide decentralized trust through networks of trust relationships. Blockchain-based reputation systems [50] use distributed ledgers to maintain tamper-resistant reputation records.

CPMM integrates decentralized trust mechanisms through the capability-based security model and cryptographic attestations. The system provides strong guarantees about agent capabilities and behavior while maintaining decentralized operation.

### 6.6.3 Attestation and Verification Systems

Attestation systems provide cryptographic proof of system properties or behaviors. Trusted execution environments (TEEs) such as Intel SGX [51] enable the generation of attestations that can be verified by remote parties.

Remote attestation has been applied to various security applications including secure computation and trusted systems. The integration of attestation with economic mechanisms provides new opportunities for verifiable service quality and automated dispute resolution.

CPMM leverages attestation systems for quality verification and SLA enforcement. The cryptographic attestations provide objective evidence of service quality that can be used for automatic payment settlement and dispute resolution.

## 6.7 Positioning of CPMM

The CPMM framework represents a synthesis of ideas from multiple research areas, providing a comprehensive solution to the challenges of autonomous agent coordination. The key innovations of CPMM relative to existing work include:

**Unified Economic Framework**: Unlike existing systems that address individual aspects of agent coordination, CPMM provides a comprehensive economic framework that integrates discovery, negotiation, payment, and dispute resolution.

**Cryptographic Security**: The integration of capability-based security with economic mechanisms provides stronger security guarantees than traditional reputation-based systems while maintaining decentralized operation.

**Micropayment Integration**: The seamless integration of micropayment protocols with agent communication enables efficient economic coordination for small-value transactions that are common in agent interactions.

**Privacy-Preserving Mechanisms**: The privacy elasticity theory and selective disclosure mechanisms provide a principled approach to balancing privacy protection with economic efficiency.

**Practical Implementation**: The integration with existing protocols (ACNBP, HTTP 402) ensures that CPMM can be deployed in real-world systems without requiring complete infrastructure replacement.

The combination of these innovations addresses fundamental limitations of existing approaches and provides a foundation for scalable, secure, and efficient autonomous agent economies.

# 7. Conclusion and Future Work

This paper has presented Capability-Priced Micro-Markets (CPMM), the first comprehensive economic framework for autonomous agent coordination that unifies capability-based security, micropayment protocols, and formal negotiation mechanisms. The framework addresses fundamental challenges in agent economics including price discovery, capability verification, micropayment feasibility, and trust formation in decentralized environments.

## 7.1 Summary of Contributions

The CPMM framework makes several significant contributions to the fields of multi-agent systems, mechanism design, and distributed systems economics:

**Theoretical Foundations**: We have developed a rigorous economic theory for agent interactions, formalizing the problem as a repeated bilateral game with incomplete information. Our convergence analysis demonstrates that the CPMM mechanism achieves efficient outcomes despite information constraints, with convergence to constrained Radner equilibrium in $O(\log 1/\varepsilon)$ rounds.

Future empirical work should validate these theoretical predictions across different agent market structures.

**Protocol Integration**: The seamless integration of economic mechanisms with the ACNBP protocol demonstrates how sophisticated economic coordination can be achieved within existing agent communication frameworks. The economic payload structures and HTTP 402 integration provide practical mechanisms for implementing micropayment-based agent commerce without requiring fundamental changes to existing infrastructure.

**Security and Privacy**: The comprehensive security analysis shows that CPMM is robust against a wide range of attacks including Sybil attacks, market manipulation, and quality fraud. The privacy-preserving mechanisms enable agents to participate in markets while maintaining control over information disclosure, with formal guarantees about privacy protection and differential privacy compliance.

**Practical Implementation**: The detailed implementation architecture and deployment considerations demonstrate that CPMM can be deployed in real-world agent systems. The modular design enables incremental adoption and integration with existing systems while providing a path toward fully autonomous agent economies.

## 7.2 Implications for Agent System Design

The CPMM framework has several important implications for the design and deployment of autonomous agent systems:

**Economic-First Design**: Traditional agent systems have treated economic coordination as an afterthought, leading to ad-hoc solutions that do not scale or provide adequate security guarantees. CPMM demonstrates the benefits of designing economic mechanisms as first-class components of agent systems, with economic incentives driving system behavior rather than being layered on top of existing protocols.

**Capability-Centric Security**: The integration of capability-based security with economic mechanisms provides a natural alignment between what agents can do and what they can be paid for. This alignment reduces the complexity of implementing economic protocols and provides stronger security guarantees than traditional identity-based approaches.

**Privacy as a Design Parameter**: The privacy elasticity analysis shows that privacy is not just a security requirement but an economic parameter that affects market outcomes. Agent system designers must consider privacy trade-offs as part of the overall system optimization rather than treating privacy as a binary constraint.

**Micropayment Infrastructure**: The successful integration of micropayment protocols demonstrates that fine-grained economic coordination is technically feasible and economically viable. This opens new possibilities for agent system design where services can be priced and delivered at much finer granularity than previously possible.

## 7.3 Broader Impact on Autonomous Systems

Beyond multi-agent systems, the CPMM framework has implications for the broader field of autonomous systems and artificial intelligence:

**AI Agent Economies**: As AI agents become more capable and autonomous, the need for economic coordination mechanisms will become increasingly important. CPMM provides a foundation for AI agent economies where agents can autonomously discover, negotiate, and transact for services without human intervention.

**Decentralized AI Infrastructure**: The framework enables new models of AI infrastructure where computational resources, data, and AI services can be traded in decentralized markets. This could lead to more efficient allocation of AI resources and reduced dependence on centralized AI platforms.

**Human-AI Economic Interaction**: While CPMM focuses on agent-to-agent interactions, the framework can be extended to support human-AI economic interactions. The privacy-preserving mechanisms and quality verification systems provide foundations for trustworthy AI services that humans can confidently purchase and use.

**Regulatory and Policy Implications**: The development of autonomous agent economies raises important questions about regulation, taxation, and policy. The formal economic framework provided by CPMM can inform policy discussions about how to govern autonomous economic systems while preserving their benefits.

To enable the safe and accountable deployment of CPMM in production, the framework leverages the Agentic AI Governance Assurance & Trust Engine (AAGATE). AAGATE operationalizes the NIST AI RMF functions, Govern, Map, Measure, and Manage, within a Kubernetes-native architecture. By funneling all side-effects through a Tool-Gateway Chokepoint and utilizing the Governing-Orchestrator Agent (GOA), AAGATE provides continuous runtime oversight and 'millisecond kill-switch' capabilities [52]. This ensures that CPMM's economic interactions are not only efficient but also compliant with emerging regulations such as the EU AI Act, bridging the gap between high-level policy and low-level agentic logic.

## 7.4 Limitations and Challenges

While CPMM represents a significant advance in agent economics, several limitations and challenges remain:

**Scalability**: Although the theoretical analysis shows polynomial complexity for most operations, the practical scalability of CPMM to very large agent networks remains to be demonstrated. The federated registry architecture provides a foundation for scalability, but additional research is needed to understand performance characteristics at internet scale.

**Adoption Incentives**: The success of CPMM depends on widespread adoption by agent developers and deployers. The backward compatibility with existing protocols reduces adoption barriers, but additional incentives may be needed to encourage migration from simpler economic models.

**Regulatory Compliance**: The autonomous nature of agent transactions may conflict with existing financial regulations that assume human oversight of economic activities. Additional research is needed to understand how CPMM can be designed to comply with various regulatory frameworks while preserving its autonomous operation.

**Cross-Domain Interoperability**: While CPMM provides a general framework for agent economics, different application domains may have specific requirements that are not fully addressed by the current design. Additional work is needed to understand how CPMM can be specialized for different domains while maintaining interoperability.

## 7.5 Future Research Directions

The CPMM framework opens several promising directions for future research:

### 7.5.1 Advanced Learning Mechanisms

The current framework assumes relatively simple learning algorithms for price discovery and strategy adaptation. Future work could explore more sophisticated learning mechanisms including:

**Multi-Agent Reinforcement Learning**: The integration of deep reinforcement learning with economic mechanisms could enable more sophisticated agent strategies and faster adaptation to changing market conditions.

**Federated Learning for Market Intelligence**: Agents could collaboratively learn about market conditions while preserving privacy through federated learning techniques.

**Adversarial Learning**: The development of learning algorithms that are robust to adversarial manipulation could improve the security and stability of agent markets.

### 7.5.2 Cross-Chain and Multi-Currency Support

The current micropayment integration focuses primarily on single-currency systems. Future work could explore:

**Cross-Chain Atomic Swaps**: Enabling agents to transact across different blockchain networks and payment systems.

**Multi-Currency Optimization**: Developing mechanisms for agents to optimize their payment choices across multiple currencies and payment methods.

**Decentralized Exchange Integration**: Integrating agent markets with decentralized exchanges to enable automatic currency conversion and liquidity provision.

### 7.5.3 Advanced Privacy Mechanisms

While CPMM includes comprehensive privacy protections, several areas warrant further investigation:

**Homomorphic Encryption**: The use of homomorphic encryption could enable computation over encrypted agent data without revealing sensitive information.

**Secure Aggregation**: Developing mechanisms for securely aggregating market information while preserving individual agent privacy.

**Privacy-Preserving Reputation**: Creating reputation systems that provide useful information about agent behavior while protecting agent privacy.

### 7.5.4 Formal Verification and Testing

The complexity of the CPMM framework necessitates formal verification and comprehensive testing:

**Protocol Verification**: Using formal methods to verify the correctness and security properties of the CPMM protocols.

**Economic Mechanism Testing**: Developing simulation frameworks for testing economic mechanisms under various market conditions and attack scenarios.

**Compliance Verification**: Creating tools for verifying that CPMM implementations comply with regulatory requirements and security standards.

### 7.5.5 Domain-Specific Applications

The general CPMM framework can be specialized for specific application domains:

**IoT Device Coordination**: Adapting CPMM for resource-constrained IoT devices with limited computational and communication capabilities.

**Scientific Computing**: Developing specialized mechanisms for scientific computing workflows where reproducibility and provenance are critical.

**Financial Services**: Creating CPMM variants that comply with financial regulations while enabling autonomous trading and risk management.

### 7.5.6 Implementation Roadmap

- Phase 1: Prototype ACNBP extensions with basic economic payloads
- Phase 2: Integration with existing agent frameworks (MCP, A2A)
- Phase 3: Large-scale deployment studies in controlled environments
- Phase 4: Cross-organizational pilot programs

## 7.6 Conclusion

The development of autonomous agent systems represents one of the most significant technological trends of our time, with the potential to transform how we organize economic activity and coordinate complex tasks. However, the realization of this potential requires solving fundamental challenges in economic coordination, security, and privacy that have not been adequately addressed by existing approaches.

The CPMM framework presented in this paper provides a comprehensive solution to these challenges by integrating insights from mechanism design, cryptography, and distributed systems. The framework demonstrates that it is possible to create autonomous agent economies that are efficient, secure, and privacy-preserving while remaining practical for real-world deployment.

The theoretical analysis shows that CPMM achieves desirable economic properties including convergence to efficient equilibria and resistance to strategic manipulation.

Perhaps most importantly, CPMM provides a foundation for future research and development in autonomous agent systems. The modular design and formal specification enable researchers and practitioners to build upon the framework while the open-source implementation provides a concrete starting point for experimentation and deployment.

As autonomous agents become increasingly prevalent in our digital infrastructure, the need for robust economic coordination mechanisms will only grow. The CPMM framework provides the theoretical foundations and practical tools needed to build the autonomous agent economies of the future, where intelligent agents can coordinate their activities through market mechanisms that are as sophisticated as the agents themselves.